\documentclass[11pt]{article}   	
\usepackage{geometry}                		
\usepackage{graphicx}				
\usepackage{amssymb}
\usepackage{array}

\usepackage{tabularx}
\usepackage{longtable}

\usepackage[affil-it]{authblk}
\usepackage{memhfixc}
\usepackage[sort&compress,numbers]{natbib}
\usepackage{color,soul}
\definecolor{darkred}{rgb}{0.9, 0.0, 0.0}
\definecolor{darkgreen}{rgb}{0.0, 0.5, 0.0}
\definecolor{darkblue}{rgb}{0.0, 0.0, 0.55}
\usepackage[pdftex,colorlinks,bookmarks,linkcolor=darkgreen, citecolor=blue, urlcolor=darkblue]{hyperref}
\usepackage[utf8]{inputenc}
\usepackage{doi}
\usepackage{bm}
\usepackage{feynmp-auto}
\usepackage{slashed}
\usepackage{amsmath}
\usepackage{tikz}
\usetikzlibrary{positioning}
\usepackage{multirow}
\usepackage{multicol}
\usepackage{booktabs}
\usepackage{eso-pic}

 \usepackage{caption}
 
 \usepackage{float}
 
 \makeatletter
\newcommand{\vast}{\bBigg@{4}}
\newcommand{\Vast}{\bBigg@{5}}
\makeatother

\newcommand{\be}{\begin{equation}}
\newcommand{\ee}{\end{equation}}
\def\ba{\begin{eqnarray}}
\def\ea{\end{eqnarray}}
\def\nn{\nonumber}
\newcommand{\nl}{\nonumber \\ }

\allowdisplaybreaks

\title{\bf General Heavy WIMP Nucleon Elastic Scattering}

\author[1,2]{Qing Chen \thanks{cqpb@ustc.edu.cn}}
\affil[1]{Interdisciplinary Center for Theoretical Study, University of Science and Technology of China, Hefei, Anhui 230026, China\vspace{1.2mm}}
\affil[2]{Peng Huanwu Center for Fundamental Theory, Hefei, Anhui 230026, China\vspace{1.2mm}}

\author[1,2]{Gui-Jun Ding \thanks{dinggj@ustc.edu.cn}}

\author[3,4]{Richard J. Hill \thanks{Richard.Hill@uky.edu}}
\affil[3]{Department of Physics and Astronomy, University of Kentucky, Lexington, KY 40506, USA \vspace{1.2mm}}
\affil[4]{Theoretical Physics Department, Fermilab, Batavia, IL 60510, USA \vspace{1.2mm}}

\begin{document}

\AddToShipoutPictureFG*{

    \AtPageUpperLeft{\put(-100,-75){\makebox[\paperwidth][r]{FERMILAB-PUB-23-423-T}}}  
     \AtPageUpperLeft{\put(-108,-60){\makebox[\paperwidth][r]{USTC-ICTS/PCFT-23-26}}}  
    }
    
\maketitle

\abstract{
Heavy WIMP (weakly-interacting-massive-particle) effective field theory is used to compute the WIMP-nucleon scattering rate for general heavy electroweak multiplets
through order $m_W/M$, where $m_W$ and $M$ denote the electroweak and WIMP mass scales. 
The lightest neutral component of such an electroweak multiplet is a candidate dark matter particle, either elementary or composite.
Existing computations for certain representations of electroweak $\mathrm{SU(2)}_W\times \mathrm{U(1)}_Y$ reveal a cancellation of amplitudes from different effective operators at leading and subleading orders in $1/M$, yielding small cross sections that are below current dark matter direct detection experimental sensitivities.
We extend those computations and consider all low-spin (spin-0, spin-1/2, spin-1, spin-3/2) heavy electroweak multiplets with arbitrary 
$\mathrm{SU(2)}_W\times \mathrm{U(1)}_Y$ representations 
and provide benchmark cross section results for dark matter direct detection experiments. 
For most self-conjugate TeV WIMPs with isospin $\le 3$, the cross sections are below current experimental limits but within reach of next-generation experiments.  An exception is the case of pure electroweak doublet, where WIMPs are hidden below the neutrino floor.}

\section{Introduction}
The field of dark matter direct detection~\cite{LZ:2022lsv,PandaX-4T:2021bab,Aprile:2018dbl,Akerib:2016vxi} comprises a large class of experiments mainly designed to detect WIMPs (Weakly-Interacting-Massive-Particles)~\cite{Lee:1977ua,Gunn:1978gr,Steigman:1978wqb,Ellis:1983ew,Goodman:1984dc,Goldberg:1983nd,Bertone:2004pz,Feng:2010gw,Bramante:2015una,Arcadi:2017kky,Roszkowski:2017nbc}. 
WIMPs can naturally explain the astronomically observed relic abundance of dark matter mass density, created in thermal equilibrium with other particles in the early Universe~\cite{Jungman:1995df}.
The primary signal process for these experiments is elastic scattering of WIMPs from atomic nuclei, detected by observing the recoiling nucleus.  Since the precise nature of the dark matter particle is unknown, the WIMP-nucleus cross section is a priori unknown.   
In order to make predictions, the problem can be approached from the ``top-down" or the ``bottom-up" perspective.  
In the ``top-down" approach, a specific UV complete theory determines all possible couplings between the new particle and Standard Model (SM) particles~\cite{Drees:1993bu}; however, the parameter space of all new physics models is large and predictions rely on model assumptions.   
In the ``bottom-up" approach, a non-relativistic expansion enforcing only the constraints of spacetime symmetries can be employed~\cite{Fan:2010gt,Fitzpatrick:2012ix}. 
While this approach does not depend on the underlying ultra-violet (UV) theory of the dark matter, 
the coefficients of the associated effective operators, and hence the dark matter-nucleus scattering rate, are undetermined.

The null results of collider searches up to a few hundred GeV~\cite{Gravili:2809396} and thermal relic abundance estimates~\cite{Cirelli:2005uq,Beneke:2020vff} suggest WIMP masses greater than the electroweak scale, 
$M \gtrsim {\rm few} \times 100\,{\rm GeV}
\gg m_W$. 
Heavy WIMP effective theory (HWET) is operative in this regime and has advantages of both the ``top-down" and ``bottom-up" approaches.   By using the scale separation between $m_W$ and $M$, HWET describes large classes of UV theories, and predicts absolute WIMP-nucleus cross sections.

The interactions of electroweak-charged 
WIMPS with quarks and gluons 
involve two classes of quark and gluon operators, 
transforming as spin-0 and spin-2, 
that largely cancel at the amplitude level, 
resulting in an ``accidentally" suppressed cross section. 
Such cancellations have been found using relativistic WIMP effective theory in specific UV completions~\cite{Hisano:2011cs}, 
and in HWET at both leading and 
subleading power~\cite{Hill:2011be,Hill:2014yka,Hill:2014yxa}. 
For example, for benchmark Wino-like or Higgsino-like particles, 
the leading order of HWET predicts a cross section one or a few orders of magnitude smaller~\cite{Hill:2013hoa} than the current experimental limits.
In fact, the cancellation essentially remains after 
including $1/M$ power corrections, and accounting for potential 
differences in nuclear responses for spin-0 and spin-2 channels~\cite{Chen:2018uqz,Chen:2019gtm}: 
subleading contributions do not lift the cross sections up to the discovery limits of current direct detection experiments.

The next generation dark matter direct detection experimental sensitivities will be improved by orders of magnitude~\cite{XENON:2015gkh,LZ:2015kxe,Liu:2017drf}, and more stringent constraints on the supersymmetric electroweak multiplets will be placed by collider experiments~\cite{ATL-PHYS-PUB-2022-018,Dainese:2019rgk}. 
To shed further light on the above-mentioned amplitude cancellation and to determine sensitivity targets for next generation experiments, 
we consider more general WIMPs with arbitrary electroweak representation and spin.
We aim to carry out a thorough survey of general electroweak-charged heavy WIMPs and compute their cross sections for scattering on a nucleon utilizing heavy WIMP effective theory through first subleading power, providing benchmark theoretical results for future direct detection phenomenology.
 Our computations will show that, for most of the self-conjugate WIMPs with low isospins, the WIMP and nucleon elastic scattering spin-independent cross sections naturally lie close to the neutrino floor of direct detection experiments, and are within striking range of next generation experiments at or below the neutrino floor. 

The remainder of the paper is organized as follows. Section.\ref{sec:hwet} constructs the heavy WIMP effective theory at the electroweak scale including order $1/M$ power corrections.
Section.\ref{sec:leet} constructs the low energy effective theory containing the WIMP and low-energy QCD.
Section.\ref{sec:match} matches the electroweak scale heavy WIMP effective theory onto this low energy effective theory.
Section.\ref{sec:uv_compeletion} illustrates minimal UV completions of the electroweak scale heavy WIMP effective theory.
Section.\ref{sec:cs} computes the cross sections for WIMP-proton elastic scattering and provides comparisons with experimental sensitivities.
Section.\ref{sec:summ} is a summary.

\section{Subleading Power Heavy WIMP Effective Theory\label{sec:hwet}}

Let us consider an electroweak multiplet with mass $M$ large compared to the weak scale, and construct the effective theory for this heavy particle in powers 
of $1/M$.
We restrict attention to the case of a self-conjugate heavy particle;
this forbids tree level $Z^0$ boson interactions, 
enabling the particle to survive current experimental exclusion limits.
Universal behavior is shared by  heavy WIMPs of different spin at the leading order of 
the heavy WIMP effective theory~\cite{Hill:2011be}.
We here investigate effects through subleading order $1/M$ and will consider 
Lorentz spin-0, spin-1/2, spin-1 and spin-3/2 WIMPs.

\subsection{Effective lagrangians}

The effective Lagrangian in the one-heavy-particle sector takes the following form for 
spin-0, spin-1/2, spin-1 and spin-3/2: 
\ba
\mathcal{L}_{\rm HWET}^{\rm spin-0}
&=&
{\phi}^\dagger_v\bigg[i v\cdot D- \delta m-\frac{{D}_\perp^2}{2M}-
\frac{f(H)}{M} - \frac{g(W,B)}{M} + \dots \bigg]\phi_v \,,
\label{H0EFT}
\\
\mathcal{L}_{\rm HWET}^{\rm spin-1/2}&=&
\bar{\chi}_v\bigg[i v\cdot D- \delta m-\frac{{D}_\perp^2}{2M}-\frac{f(H)}{M}
- \frac{g(W,B)}{M}
+ \dots \bigg]\chi_v \,,
\label{H12EFT}
\\
\mathcal{L}_{\rm HWET}^{\rm spin-1}&=&
\mathcal{V}^{\mu \dagger}_v\bigg[\left(i v\cdot D- \delta m-\frac{{D}_\perp^2}{2M}-\frac{f(H)}{M}\right)\left(-g_{\mu\nu}\right)
+ \frac{g(W,B)_{\mu\nu}}{M}
+ \dots \bigg]\mathcal{V}^\nu_v \,,
\label{H1EFT}
\\
\mathcal{L}_{\rm HWET}^{\rm spin-3/2}&=&
\bar{\xi}^{\mu}_v\bigg[\left(i v\cdot D- \delta m-\frac{{D}_\perp^2}{2M}-\frac{f(H)}{M}\right)\left(-g_{\mu\nu}\right)
+ \frac{g(W,B)_{\mu\nu}}{M}
+ \dots \bigg]\xi^\nu_v \,.
\label{H32EFT}
\ea
Here the ellipses denote terms of order $1/M^2$, $v^\mu$ is the heavy WIMP velocity with $v^2=1$, and $\delta m$ is a residual mass matrix after integrating out the heavy particle.  The covariant derivative is $D_\mu=\partial_\mu- i g_1 Y B_\mu-i g_2 W_\mu^a T^a$
where $Y$ is the $\mathrm{U(1)}_Y$ hypercharge and $T^a$ are $\mathrm{SU(2)}_W$ generators, with $a=1,2,3$.
Perpendicular components are 
projected using $g_\perp^{\mu\nu}\equiv g^{\mu\nu}-v^\mu v^\nu$ as usual
(thus e.g. $ \sigma^{\mu\nu}_{\perp}= g_\perp^{\alpha\mu} g_\perp^{\beta \nu} \sigma_{\alpha\beta}$,  
 $D_\perp^\mu = D^\mu - v^\mu v\cdot  D$). 
 We have applied field redefinitions to remove redundant operators, and enforced the constraints for expressing
 heavy particles in terms of four-component Dirac spinors with vector indices (e.g. 
 $v_\mu\mathcal{V} ^\mu_v=0$, $\slashed{v} \xi^\mu_v=\xi^\mu_v$
  and $\gamma_\mu \xi^\mu_v=0$)~\cite{Heinonen:2012km}.
The Higgs-WIMP interaction $f(H)$ will be discussed below.  Terms contained in $g(W,B)$ give rise to suppressed spin-dependent scattering rates and will not be considered further in this work~\cite{Chen:2019gtm}. 

\subsection{Higgs interactions}

The Higgs-WIMP interaction, $f(H)$, depends on the specific spin and electroweak representation of the WIMP.
Let us construct gauge- and Lorentz- invariant operators containing WIMP and Higgs fields, for a general $(J,\, Y)$ representation of ${\rm SU}(2)_W\times {\rm U}(1)_Y$,  with $J$ being ${\rm SU}(2)$ isospin and $Y$ being ${\rm U}(1)$ hypercharge. 
Since we are concerned with the one-WIMP sector, there must be two WIMP fields in the Lagrangian interaction.
Since the Standard Model Higgs $H$ is a $(1/2, 1/2)$ representation under  ${\mathrm {SU}}(2)_{ W}\times { \mathrm{U}}(1)_{\mathrm Y}$,  there is no three-point gauge-invariant interaction.
The leading WIMP-Higgs interaction arises from 
four-point interactions with two WIMP fields and two Higgs fields.    
The Higgs bilinear $H^\dagger H$ transforms as a singlet, 
and $H^\dagger \tau^a H$ transforms as a triplet with $\tau^a$ the isospin Pauli matrices.
It is convenient to introduce $\tilde{H}\equiv i\tau^2 H^*$, 
which transforms identically to $H$ under ${\mathrm {SU}}(2)$ 
but has opposite hypercharge.

The possible forms for 
$f(H)$ can be tabulated by first considering manifestly Lorentz-invariant Lagrangians and then making the identifications
\begin{align}
&\mathrm{spin-0: }\hspace{10pt}
\Phi(x)=\sqrt{\frac{1}{M}}e^{-iMv\cdot x}\phi_v(x) 
\,,\label{fr:scalar}
\\
&\mathrm{spin-1/2: }\hspace{10pt}
\chi=\sqrt{2}e^{-iMv\cdot x}\left(\chi_v+X_v\right)\,,\label{fr:majorana}
\\
&\mathrm{spin-1: }\hspace{10pt}
\mathcal{V}^\mu(x)=\sqrt{\frac{1}{M}}e^{-iMv\cdot x}
\mathcal{V}^\mu_v(x) \,,\label{fr:vec}
\\
&\mathrm{spin-3/2: }\hspace{10pt}
\xi^\mu=\sqrt{2}e^{-iMv\cdot x}\left(\xi_v^\mu+\Xi_v^\mu\right)\,,
\label{fr:RS}
\end{align}
where for $\chi$ and $\xi^\mu$, the second terms in parentheses denote anti-particle degrees of freedom that are integrated out
(the components satisfy $\slashed{v} \chi_v = \chi_v$, $\slashed v \xi^\mu = \xi^\mu$, and $\slashed{v} X_v = -X_v$, $\slashed{v} \Xi^\mu = -\Xi^\mu$). 
Let us consider separately the cases of spin 0, $1/2$, $1$ and $3/2$. 

\subsubsection{Spin 0}

The relativistic spin-0 electroweak multiplet and Higgs interaction takes the form
\begin{align}
\mathcal{L}_H^{\rm spin-0}&=c_{1,0}\phi^\dagger\phi H^\dagger H+c_{2,0}\phi^\dagger t^a \phi H^\dagger \tau^a H
\nl
&+\left(c_{3,0}\phi^\dagger t^a \tilde{\phi} \tilde{H}^\dagger \tau^a H+\rm{ h.c.} \right)\delta_{Y,\, 1/2}
\nl
&+\left(c_{4,0}\phi^\dagger t^a \tilde{\phi} H^\dagger \tau^a \tilde{H}+\rm{h.c.} \right)\delta_{Y,\,-1/2}\,,
\label{spin0H}
\end{align}
where $t^a$ is an ${\mathrm {SU}}(2)$ generator, $\tilde{\phi}\equiv U\phi^*$ and $U$ is a $(2J+1)\times (2J+1)$ matrix acting as a similarity transformation for isospin-$J$ representation \cite{Chao:2018xwz},
\be
U\left(e^{i\boldsymbol{\alpha\cdot t}}\right)^*U^{-1}=e^{i\boldsymbol{\alpha\cdot t}}\,,
\ee
where $\boldsymbol{t}=(t^1, t^2, t^3)$ are ${\rm SU}(2)$ generators and $\boldsymbol{\alpha}=(\alpha_1 ,\alpha_2, \alpha_3)$ are real parameters.  The explicit matrix elements are
\begin{align}
&t^1_{mn}=\left(\sqrt{m(2J+1-m)}\delta_{n, m+1}+\sqrt{n(2J+1-n)}\delta_{n, m-1}\right)/2\,, 
\nl
&t^2_{mn}=-i\left(\sqrt{m(2J+1-m)}\delta_{n, m+1}-\sqrt{n(2J+1-n)}\delta_{n, m-1}\right)/2\,,
\nl
&t^3_{mn}=(J+1-m)\delta_{mn} \,,
\nl
&U_{mn}=(-1)^{m+1}\delta_{m+n,2J+2}\,,
\end{align}
where $m,\, n=1,\,2,\,...,\, 2J+1$.
In particular, when $J=1/2$, $U$ is the matrix $i\tau^2$ that we have introduced above in the construction of $\tilde{H}$.
The coefficient $c_{i,0}$ has a subscript $0$ standing for spin-0.
The notation h.c. denotes hermitian conjugate.
Gauge-invariant interactions among electroweak multiplets including Higgs field can also be obtained by brute-force construction of gauge-singlets using Clebsch-Gordan coefficients \cite{Chao:2018xwz}, and we have checked the equivalency of the two methods.

When the hypercharge $Y=0$, we restrict attention to integer isospin, for which there is an electrically neutral dark matter candidate.   For integer isospin, the ${\mathrm {SU}}(2)$ representation is real, and the scalar field $\phi$ 
may be chosen real, and identified with $\Phi$ in Eq.~(\ref{fr:scalar}).  
Interaction $c_{2,0}$ vanishes in this case, and only $c_{1,0}$ appears;  according to Eq.~(\ref{fr:scalar}),  
$f(H)=c_{1,0}H^\dagger H \equiv -c_H H^\dagger H$~\cite{Hill:2011be,Chen:2018uqz}. 
When $Y\neq 0$, the field $\Phi$ in Eq.~(\ref{fr:scalar}) 
is identified with the column vector of two real scalar fields, 
\begin{align}
\Phi &= \left(\begin{array}{c} \frac{1}{\sqrt{2}}(\phi+\phi^*) \\
\frac{i}{\sqrt{2}}(\phi-\phi^*)
\end{array}\right) \,, 
\end{align}
and $f(H)$ may be read off according to Eq.~(\ref{fr:scalar}).

\subsubsection{Spin 1/2}

For a spin-1/2 electroweak multiplet,
let us construct the self-conjugate fields from Weyl spinors, 
$\psi_L$ and $\psi_L^\prime$, 
transforming under ${\mathrm {SU}}(2)_{ W}\times {\mathrm {U}}(1)_{Y}$ as
\ba
&&\psi_L\rightarrow e^{i\boldsymbol{\alpha\cdot t}}e^{i\beta Y}\psi_L\,,\nl
&&\psi_L^\prime\rightarrow e^{-i\boldsymbol{\alpha\cdot t^*}}e^{-i\beta Y}\psi_L^\prime\,.
\label{su2trans}
\ea
The general spin-1/2 Higgs interaction is then
\begin{align}
\mathcal{L}_H^{\rm spin-1/2}&=
-\frac{1}{M}\Bigg[c_{1,\frac{1}{2}}H^\dagger H \left({\psi}^{\prime T}_L i\sigma^2\psi_L\right)+c_{2,\frac{1}{2}}H^\dagger \tau^a  H \left({\psi}^{\prime T}_L i\sigma^2 t^a \psi_L\right)
\nl
&+c_{3,\frac{1}{2}}{H}^\dagger \tau^{a } \tilde{ H} \left({\psi}^{T}_L i\sigma^2 U^\dagger t^{a } \psi_L\right)\delta_{Y,\,1/2}
+c_{4,\frac{1}{2}}\tilde{H}^\dagger \tau^a { H} \left({\psi}^{\prime T}_L i\sigma^2 t^a U\psi_L^\prime\right)\delta_{Y,\,1/2}
\nl
&+c_{5,\frac{1}{2}}\tilde{H}^\dagger \tau^{a} { H} \left({\psi}^{T}_L i\sigma^2 U^\dagger t^{a } \psi_L\right)\delta_{Y,\,-1/2}
+c_{6,\frac{1}{2}}H^\dagger \tau^a \tilde{ H} \left({\psi}^{\prime T}_L i\sigma^2 t^a U\psi_L^\prime\right)\delta_{Y,\,-1/2}
+\rm{h.c.}\Bigg]\,.
\label{12H}
\end{align}

To connect with Eq.~(\ref{fr:majorana}), let us embed $\psi_L$ and $\psi_L^\prime$ into the Dirac field $\psi$, with its conjugate $\psi^c$:
\ba\label{eq:psidef}
\psi=
\begin{pmatrix}
          \psi_L\\
           i\sigma^2\psi_L^{\prime *} \\
\end{pmatrix}, 
\hspace{1cm}
\psi^c=
\begin{pmatrix}
          \psi_L^{\prime}\\
           i\sigma^2\psi_L^{*} \\
\end{pmatrix}\,,
\ea
where the Pauli matrix $\sigma^2$ acts as a generator in the Lorentz group.
The self-conjugate (Majorana) fermion field $\chi$ in Eq.~(\ref{fr:majorana}) 
is then identified with 
\begin{align}\label{eq:chidef}
    \chi &= \left( \begin{array}{c} \chi_1 \\ \chi_2 \end{array}\right) 
    = \left(\begin{array}{c} \frac{1}{\sqrt{2}}(\psi+\psi^c) \\
\frac{i}{\sqrt{2}}(\psi-\psi^c)
\end{array}
\right)
    \,.
\end{align}

For hypercharge $Y=0$ and integer isospin, we may choose 
irreducible representations involving a single Weyl fermion, 
i.e. $\psi_L^\prime =\psi_L$.  For this case, all interactions 
except $c_{1,\frac12}$ vanish.  
Coefficient $c_{1,\frac12}$ may be chosen real%
\footnote{
This may be obtained by field redefinition $\delta \psi_L \sim (H^\dagger H/M^2)\psi_L$. 
}
and according to Eq.~(\ref{fr:majorana}), $f(H)=c_{1,\frac12} H^\dagger H \equiv -2 c_H H^\dagger H$~\cite{Chen:2019gtm}.
For the Dirac fermion (``Higgsino") case, $J=Y=1/2$, the gauge-invariant interaction (\ref{12H})  may be simplified.  After expressing $\psi_L$ and $\psi_L^\prime$ in terms of 
$\chi_v$ via Eqs.~(\ref{eq:psidef}), (\ref{eq:chidef})
and (\ref{fr:majorana}), 
$f(H)$ in Eq.~(\ref{H12EFT}) is expressed as a matrix with four real parameters~\cite{Chen:2019gtm}.
\footnote{The correspondence with Eq.~(2) of Ref.~\cite{Chen:2019gtm} is $a=-{\rm Re}(c_2)/2$, $b=-(c_3^*+c_4)/2$ and $c=-{\rm Re}(c_1-c_2/2)/2$.}

\subsubsection{Spin 1}

Similar to the spin-0 case, the spin-1 electroweak multiplet and Higgs bilinear interaction takes the form
\begin{align}
\mathcal{L}_H^{\rm spin-1}&=c_{1,1}V^{\mu \dagger}V_\mu H^\dagger H+c_{2,1}V^{\mu \dagger}t^aV_\mu H^\dagger \tau^a H
+\left(c_{3,1}V^{\mu \dagger} t^a \tilde{V}_\mu  \tilde{H}^\dagger \tau^a H+\rm{ h.c.} \right)\delta_{Y,1/2}
\nl
&+\left(c_{4,1}V^{\mu \dagger}  t^a  \tilde{V}_\mu  H^\dagger \tau^a \tilde{H}+\rm{h.c.} \right)\delta_{Y,-1/2}\,,
\label{vH}
\end{align}
where $\tilde{V}_\mu=UV_\mu^*$, and the self-conjugate basis consists of two real vectors,
\ba
V^\mu_1=\frac{1}{\sqrt{2}}(V^\mu+V^{\mu\dagger}),
\hspace{1cm}
V^\mu_2=\frac{i}{\sqrt{2}}(V^\mu-V^{\mu\dagger})\,,
\ea
where $\mathcal{V}^\mu=\left(V^\mu_1, V^\mu_2\right)^T$ is the relativistic field mapping onto the heavy vector $\mathcal{V}_v^\mu$ in  Eq.\,(\ref{fr:vec}).

\subsubsection{Spin 3/2}

Similar to the spin-1/2 case, the spin-3/2 electroweak multiplet and Higgs bilinear interaction takes the form
\begin{align}
\mathcal{L}_H^{\rm spin-3/2}&=
\frac{1}{M}\Bigg[c_{1,\frac{3}{2}}H^\dagger H \left({\varPsi}^{\prime \mu  T}_L i\sigma^2\varPsi_{\mu L}\right)
+c_{2,\frac{3}{2}}H^\dagger \tau^a  H \left({\varPsi}^{\prime \mu  T}_L i\sigma^2 t^a \varPsi_{\mu L}\right)
\nl
&+c_{3,\frac{3}{2}}{H}^\dagger \tau^{a } \tilde{ H} \left({\varPsi}^{ \mu  T}_L i\sigma^2 U^\dagger t^{a } \varPsi_{\mu L}\right)\delta_{Y,\,1/2}
+c_{4,\frac{3}{2}}\tilde{H}^\dagger \tau^a { H} \left({\varPsi}^{\prime \mu  T}_L i\sigma^2 t^a U \varPsi_{\mu L}^\prime\right)\delta_{Y,\,1/2}
\nl
&+c_{5,\frac{3}{2}}\tilde{H}^\dagger \tau^{a} { H} \left({\varPsi}^{ \mu  T}_L i\sigma^2 U^\dagger t^{a } \varPsi_{\mu L} \right)\delta_{Y,\,-1/2}
+c_{6,\frac{3}{2}}H^\dagger \tau^a \tilde{ H} \left({\varPsi}^{\prime \mu  T}_L i\sigma^2 t^a U \varPsi_{\mu L}^\prime\right)\delta_{Y,\,-1/2}
+\rm{h.c.}\Bigg]\,,
\label{RSH}
\end{align}
where we have the Rarita-Schwinger field $\varPsi^\mu$ and its conjugate $\varPsi^{c\, \mu}$,
\ba
\varPsi^\mu=
\begin{pmatrix}
         \varPsi_L^\mu\\
           i\sigma^2\varPsi_L^{\prime \mu *} \\
\end{pmatrix}, 
\hspace{1cm}
\varPsi^{\mu\, c}=
\begin{pmatrix}
           \varPsi_L^{\prime \mu}\\
           i\sigma^2  \varPsi_L^{\mu *} \\
\end{pmatrix}\,.
\ea
Constructing self-conjugate fermions from  
$\varPsi^\mu$ and  $\varPsi^{c\, \mu}$,  
\ba
\xi^\mu_1=\frac{1}{\sqrt{2}}(\varPsi^\mu+\varPsi^{\mu \,c}),
\hspace{1cm}
\xi^\mu_2=\frac{i}{\sqrt{2}}(\varPsi^{\mu}-\varPsi^{\mu \,c})\,,
\ea
we identify  $\xi^\mu\equiv\begin{pmatrix} \xi^\mu_1, & \xi^\mu_2 \end{pmatrix}^T$ as the relativistic field in Eq.\,(\ref{fr:RS}).

\subsubsection{EWSB and Feynman rules}

After electroweak symmetry breaking, the Higgs field acquires its vacuum expectation value
\ba
\langle H\rangle=\frac{v}{\sqrt{2}}
\begin{pmatrix}
0\\
1
\end{pmatrix}\,,
\label{higgsvev}
\ea
and in the self-conjugate basis, the mass matrix becomes 
\ba
&&\delta M(v)=
\delta m+\frac{v^2}{2M}
\begin{pmatrix}
M_1+\mathbf{Re}\left(M_2\right)& -\mathbf{Im}\left(M_2\right)\\
-\mathbf{Im}\left(M_2\right)& M_1-\mathbf{Re}\left(M_2\right)
\end{pmatrix}\,,
\label{mmatrix}
\ea
where the matrices $M_1$ and $M_2$ are
\begin{align}
\left(M_1\right)_{ kl}&=(A+Bk)\delta_{kl}\,,
\nl
\left(M_2\right)_{ kl}&=\delta_{Y,\,1/2}C(-1)^{k} \sqrt{(k-1)(2J+2-k)}\delta_{k+l,\, 2J+3}
+\delta_{Y,\,-1/2}C^\prime(-1)^{k} \sqrt{k(2J+1-k)}\delta_{k+l,\, 2J+1}\,,
\end{align}
with $k,\, l=1,\,2,\,...,\,2J+1$ and 
\begin{align}
A&=\frac{1}{2}\left[\mathbf{Re}(c_{1,s})-\mathbf{Re}(c_{2,s})(J+1)\right]\,,
\nl
B&=\frac{1}{2}\mathbf{Re}(c_{2,s})\,,
\nl
C&=\frac{c^*_{3,s}+c_{4,s}}{2}\left(\delta_{s, \frac{1}{2}}+\delta_{s, \frac{3}{2}}\right)+c_{3,s}\left(\delta_{s, 0}+\delta_{s,1}\right)\,,
\nl
C^\prime&=\frac{c^*_{5,s}+c_{6,s}}{2}\left(\delta_{s, \frac{1}{2}}+\delta_{s, \frac{3}{2}}\right)+c_{4,s}\left(\delta_{s, 0}+\delta_{s,1}\right)\,,
\label{ABCexprs}
\end{align}
where $s=0, \,\frac{1}{2}, \,1, \,\frac{3}{2}$ is the spin.

Diagonalizing the mass matrix Eq.\,(\ref{mmatrix}), when isospin $J$ is half-integer (even-dimension representation) and hypercharge $Y=1/2$, the eigenvalues are
\begin{align}
&\frac{v^2}{2M}\left(A+B\right),\,\frac{v^2}{2M}\left(A+B\right)\,,
\nl
&\frac{v^2}{2M}\left[A+\left(J+\frac{3}{2}\right)B\pm\left(J+\frac{1}{2}\right)|C|\right]\,,
\nl
&\frac{v^2}{2M}\left[A+\left(J+\frac{3}{2}\right)B\pm\sqrt{n^2B^2+\left[\left(J+\frac{1}{2}\right)^2-n^2\right]|C|^2}\right]\,,
\nl
&\frac{v^2}{2M}\left[A+\left(J+\frac{3}{2}\right)B\pm\sqrt{n^2B^2+\left[\left(J+\frac{1}{2}\right)^2-n^2\right]|C|^2}\right]\,,
\end{align}
for integer $n=1,..., J-1/2$.
The two mass eigenvalues in the first row correspond to the components with the
largest magnitudes of charge, $Q=\pm (J+1/2)$.
The non-vanishing $C$ term acts to split the degenerate mass of two neutral components and 
$\frac{v^2}{2M}\left[A+\left(J+\frac{3}{2}\right)B-\left(J+\frac{1}{2}\right)|C|\right]$ in the second row is the mass of the lightest neutral component, the WIMP.
The four mass eigenvalues in the last two rows are for the two pairs of components with opposite charges $Q=\pm n$.

When isospin $J$ is a half-integer and hypercharge $Y=-1/2$, the mass eigenvalues are
\begin{align}
&\frac{v^2}{2M}\left(A+(2J+1)B\right),\,\frac{v^2}{2M}\left(A+(2J+1)B\right)\,,
\nl
&\frac{v^2}{2M}\left[A+\left(J+\frac{1}{2}\right)B\pm\left(J+\frac{1}{2}\right)|C^\prime|\right]\,,
\nl
&\frac{v^2}{2M}\left[A+\left(J+\frac{1}{2}\right)B\pm\sqrt{n^2B^2+\left[\left(J+\frac{1}{2}\right)^2-n^2\right]|C^\prime|^2}\right]\,,
\nl
&\frac{v^2}{2M}\left[A+\left(J+\frac{1}{2}\right)B\pm\sqrt{n^2B^2+\left[\left(J+\frac{1}{2}\right)^2-n^2\right]|C^\prime|^2}\right]\,,
\end{align}
with similar notation as for the $Y=1/2$ case. Here $\frac{v^2}{2M}\left[A+\left(J+\frac{1}{2}\right)B-\left(J+\frac{1}{2}\right)|C^\prime|\right]$ is the mass of the lightest neutral component, the WIMP.

When $J$ is an integer (odd-dimension representation), the hypercharge must be an integer to provide a neutral component for the WIMP.
Further, if the electroweak multiplet is self-conjugate, hypercharge must be zero.
 After electroweak symmetry breaking, the mass matrix becomes 
\ba
&\delta M(v)=
\delta m+\frac{v^2}{2M}M_1^\prime\,,
\ea
where
\ba
\left(M_1^\prime\right)_{ kl}=(A^\prime+B^\prime k)\delta_{kl},
\ea
for $k, l=1,\,2,...,\, 2J+1$, with 
\begin{align}
A^\prime&=\mathbf{Re}(c_{1,s})-\mathbf{Re}(c_{2,s})(J+1)\,,
\nl
B^\prime&=\mathbf{Re}(c_{2,s}). 
\end{align}
The eigenvalue for the neutral state is $v^2[A^\prime+B^\prime (J+1)]/(2M)$.

\begin{figure}[t]
\centering
\includegraphics[width=.2\linewidth]{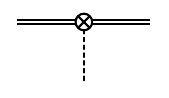}
\captionsetup{width=.9\linewidth}
\caption{
WIMP-WIMP-Higgs boson effective interaction at order $1/M$. The double line denotes the heavy WIMP field and the dashed line denotes the Higgs boson.
}
\label{3pt}
\end{figure}

After electroweak symmetry breaking, the WIMP interacts with the dynamical components of the Higgs field, 
$\phi_1$, $\phi_2$, $\phi_3$ and $h$  as follows,
\ba
H=\langle H\rangle
+
\frac{1}{\sqrt{2}}
\begin{pmatrix}
\phi_1+i\phi_2\\
h+i\phi_3
\end{pmatrix}\,,
\label{Hfield}
\ea
and 
the Lagrangian of the heavy field after electroweak symmetry breaking in the charge and mass eigenstate basis $\chi_0^{(\mu)}$ is
\begin{align}
\mathcal{L}&=
\bar{\chi}_0^{(\mu)}\Big[i v\cdot \partial+eQv\cdot A+\frac{g_2}{\cos{\theta_W}}v\cdot Z(\tilde{T}^3-\sin^2{\theta_W}Q)
+\frac{g_2}{\sqrt{2}}v\cdot(W^+\tilde{T}^++W^-\tilde{T}^-)\nn\\
&\quad- \delta M
-\frac{{D}_\perp^2}{2M}+
\frac{f(h)}{M}+\frac{g(W,\,Z)}{M}+...\Big]\chi_{0{(\mu)}}\,,
\label{brokenL}
\end{align}
where the WIMP is the lightest neutral state of the electroweak multiplet which we denote as $\chi_0^{(\mu)}$ (with or without the vector index $\mu$ depending on its spin), and the tilded gauge generator matrices refer to this basis.
Here $f(h)$ describes the WIMP interaction with 
the Higgs boson, cf. Fig.~\ref{3pt}.
The Feynman rule for this vertex, $i v c_H (g_{\mu\nu}) /M$, 
is related to the mass eigenvalue of the lightest neutral state.
When hypercharge $Y=1/2$, it is
\begin{align}
-i\frac{v}{M}\left[A+\left(J+\frac{3}{2}\right)B-\left(J+\frac{1}{2}\right)|C|\right]\,.
\label{3ptvtx_even_pos}
\end{align}
When  hypercharge $Y=-1/2$, it is
\begin{align}
-i\frac{v}{M}\left[A+\left(J+\frac{1}{2}\right)B-\left(J+\frac{1}{2}\right)|C^\prime|\right]\,.
\label{3ptvtx_even_neg}
\end{align}
When  hypercharge $Y=0$, it is
\begin{align}
-i\frac{v}{M}\left[A^\prime +\left(J+1\right)B^\prime \right]\,.
\label{3ptvtx_odd}
\end{align}

If inelastic scattering is considered, 
the relevant operators will involve components in the electroweak multiplet other than the WIMP, as well as $\phi_1$, $\phi_2$, $\phi_3$ components in the Higgs field, 
which can be found by inserting Eq.\,(\ref{Hfield}) into the $f(H)$ term in the Lagrangians, Eq.\,(\ref{H0EFT}), Eq.\,(\ref{H12EFT}), Eq.\,(\ref{H1EFT}) and 
Eq.\,(\ref{H32EFT}), then diagonalizing to the charge and mass eigenstate basis as in the above procedure. We focus on elastic scattering in this paper and leave the inelastic case to future work.

\section{Low Energy Effective Theory \label{sec:leet}}

After electroweak symmetry breaking, the dark matter particle $\chi_0^{(\mu)}$ is a singlet under $\mathrm{SU}(3)_c \times \mathrm{ U}(1)_{\rm e.m.}$.
The low energy effective operators for dark matter (DM) and nucleon scattering at the quark level can be constructed from heavy WIMP bilinears, and quark and gluon bilinears.

We focus on WIMP-nucleon spin-independent elastic scattering, which is the dominant process 
for many dark matter direct detection experiments. 
The relevant low energy effective theory for 
the spin-independent interaction of spin-0 and spin-1/2 heavy WIMP with quarks (top-quark been integrated out) and gluons is~\cite{Hill:2014yxa}
\begin{align}
\mathcal{L}=\bar{\chi}_0 \chi_0 \bigg \{\sum_{q=u,d,s,c,b}\left[c_q^{(0)}O_q^{(0)}+c_q^{(2)}v_{\mu}v_{\nu}O_q^{(2)\mu\nu} \right]+
c_g^{(0)}O_g^{(0)}+c_g^{(2)}v_{\mu}v_{\nu}O_g^{(2)\mu\nu} \bigg\} \,,
\label{5fQCD}
\end{align}
where 
\ba
&&O_q^{(0)}=m_q \bar{q}{q},\quad O^{(2) \mu \nu }_{q} = \frac{1}{2}\bar{q}\left(\gamma^{\{\mu}iD_{-}^{\nu\}}-\frac{g^{\mu \nu}}{d}i\slashed{D}_{-}\right)q  \,,
\nl
&&O^{(0)}_g=G^{A\, \mu\nu}G^A_{\mu\nu},\quad O^{(2) \mu \nu }_{g}= -G^{A\mu\lambda}G^{A\nu}_{~~~\lambda}+\frac{1}{d}g^{\mu\nu}(G^{A}_{\alpha\beta})^{2} \,,
\ea
with $D^\mu_{-} \equiv D ^\mu- \overleftarrow{D}^\mu$.

For higher-spin particles, no essentially new operators appear in 
the spin-independent sector of the low energy effective theory.
\footnote{This may be seen by constructing an explicit basis, and enforcing the constraints on higher-spin representations for heavy particles~\cite{Heinonen:2012km}, e.g. 
$v_\mu \chi_0^{\mu}=0$,
$\slashed{v} \chi_0^{\mu}= \chi_0^{\mu}$,
$\epsilon_{\nu\alpha\beta\mu}v^{\nu}\sigma^{\alpha\beta} \chi_0^{\mu}=0$.
}
For spin-1 and spin-3/2 heavy WIMPs interacting with quarks and gluons, we have 
\begin{align}
\mathcal{L}=\bar{\chi}_0^{\mu}\chi_0^{\nu} \bigg \{\sum_{q=u,d,s,c,b}\left[c_q^{(0)}O_q^{(0)}
+c_q^{(2)}v_{\alpha}v_{\beta}O_q^{(2)\alpha\beta} \right]
+c_g^{(0)}O_g^{(0)}+c_g^{(2)}v_{\alpha}v_{\beta}O_g^{(2)\alpha\beta} \bigg\}g_{\mu\nu} \,.
\label{5fQCD_hs}
\end{align}

\section{Weak Matching \label{sec:match}}

To determine the Wilson coefficients in the effective theories Eq.\,(\ref{5fQCD}) and Eq.\,(\ref{5fQCD_hs}) and obtain results for WIMP-nucleon elastic scattering, 
we match the low energy Lagrangian to the electroweak scale effective theory (\ref{brokenL}) by integrating out the weak scale particles.

\begin{figure}[t]
\begin{center}
        \captionsetup{width=.9\linewidth}
\includegraphics[width=.7\linewidth]{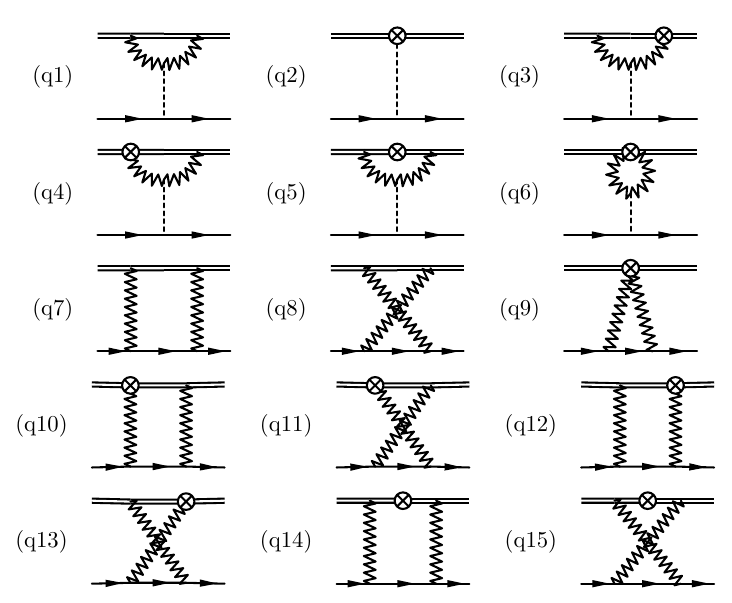}
\caption{\label{fig:1Mquarkmatching}
\small
  Diagrams contributing through $1/M$ order to quark matching.  Double lines denote the WIMP,
   solid lines denote quarks,
  zigzag lines denote weak gauge bosons, and dashed lines denote the Higgs boson.  
  The encircled cross denotes an insertion of $1/M$ order effective operators.
}
\end{center}
\end{figure}

\begin{figure}[h]
\begin{center}
        \captionsetup{width=.9\linewidth}
\includegraphics[width=.8\linewidth]{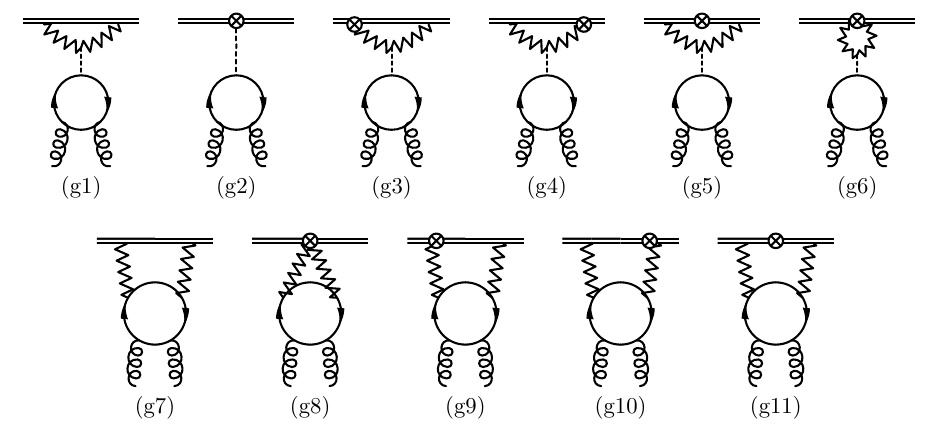}
\caption{\label{fig:gluonmatching}
\small
  Diagrams contributing through $1/M$ order to gluon matching, with the same notation as in 
  Fig.~\ref{fig:1Mquarkmatching}. Curly lines denote gluons. Diagrams with both gluons attached to the upper quark line or with one gluon attached to each of the upper and lower quark lines are not shown in the second row.
}
\end{center}
\end{figure}

The matching diagrams for WIMP and quark operators are shown in Fig.\,\ref{fig:1Mquarkmatching}. 
Note that all diagrams involving Nambu-Goldstone bosons are suppressed compared to the diagrams present in Fig.\,\ref{fig:1Mquarkmatching}.
All Standard Model particles are treated as massless except the weak scale particles  $W^\pm$, $Z^0$, $h$, $t$-quark, which will be integrated out.
The matching for WIMP and gluon operators are shown in Fig.\,\ref{fig:gluonmatching}.
The details of the matching can be found in \cite{Chen:2021tcw}.

The renormalized Wilson coefficients  in Eq.\,(\ref{5fQCD}) and Eq.\,(\ref{5fQCD_hs}) for the low energy five-flavor quarks and gluons effective theory are
\begin{align}\label{eq:results}
  \hat{c}_{U}^{(0)}(\mu)&= -\frac{1}{2 x_{h}^{2}} \left(f_W+\frac{f_Z}{c_W^3}\right) +\frac{f_Z}{8c_W}(c_V^{(U)2}-c_A^{(U)2})- \frac{m_W}{\pi M}\frac{{c}_H}{\alpha_{2}^{2} x_h^2}  \,,
  \nl
  \hat{c}_{D}^{(0)}(\mu)&= -\frac{1}{2 x_{h}^{2}} \left(f_W+\frac{f_Z}{c_W^3}\right) +\frac{f_Z}{8 c_W}(c_V^{(D)2}-c_A^{(D)2})
  -\delta_{Db}f_W \frac{x_{t}}{8(x_{t}+1)^{3}} 
  - \frac{m_W}{\pi M} \frac{{c}_H}{\alpha_{2}^{2} x_h^2}\,,
  \nl
 \hat{c}_{g}^{(0)}(\mu)&=    \frac{\alpha_s(\mu)}{4\pi}\bigg\{ \frac{1}{6}\bigg[\frac{1}{x_h^2}\Big(f_W+\frac{f_Z}{c_W^3}\Big)
  +f_W\left[\frac{N_l}{2}+\frac{1}{2(x_t+1)^2}\right]\bigg] + \frac{m_W}{ \pi M}\frac{{c}_H}{3 \alpha_2^2 x_h^2} 
  \nl
&\quad  +\frac{f_Z}{16 c_W}\bigg[(c_V^{(D)2}+c_A^{(D)2}) 
 +\frac{1}{4}(c_V^{(U)2}+c_A^{(U)2})\Big[\frac{4(48y_t^6-2y_t^4+9y_t^2-1)}{3(4y_t^2-1)^3}
 \nl
& \quad+\frac{8}{3} -\pi y_t +\frac{32y_t^6(8y_t^2-7)}{(4y_t^2-1)^{7/2}}\arctan(\sqrt{4y_t^2-1})\Big] 
 +\frac{1}{4}(c_V^{(U)2}-c_A^{(U)2})\Big[3\pi y_t 
 \nl
& \quad -\frac{32y_t^4(24y_t^4-21y_t^2+5)}{(4y_t^2-1)^{7/2}}\arctan(\sqrt{4y_t^2-1}) 
-\frac{4(144y_t^6-70y_t^4+9y_t^2-2)}{3(4y_t^2-1)^3} \Big]\bigg] 
\bigg\}
  \,,
  \nl
  \hat{c}_{U}^{(2)}(\mu)&=\frac{f_W}{3} 
+\frac{f_Z}{6 c_W}(c_V^{(U)2}+c_A^{(U)2})
 -\frac{f_Z}{4 c_W^2}(c_V^{(U)2}+c_A^{(U)2})\frac{m_W}{\pi M}
-f_W\frac{m_W}{2\pi M}  \,,
  \nl
  \hat{c}_{D}^{(2)}(\mu)&=\frac{f_W}{3}  
+\frac{f_Z}{6 c_W}(c_V^{(D)2}+c_A^{(D)2})
 -\frac{f_Z}{4 c_W^2}(c_V^{(D)2}+c_A^{(D)2})\frac{m_W}{\pi M}
-f_W\frac{m_W}{2\pi M} \nl
&\quad +f_W\frac{\delta_{Db}}{2}
\Big[ \frac{3x_t+2}{3(x_t+1)^3}-\frac{2}{3}-
\frac{m_W}{\pi M}\frac{x_t^2(1-x_t^4+4x_t^2\log{x_t})}{(x_t^2-1)^3} \Big]  \,,
  \nl
  \hat{c}_{g}^{(2)}(\mu)&=
  \frac{\alpha_s(\mu)}{4\pi}\bigg\{
 2f_W \bigg[N_\ell\left( - {4\over 9} \log{\mu\over m_W} - {1\over 2} \right) - {(2+ 3x_t)\over 9(1+x_t)^3}\log{\mu \over m_W(1+x_t)}
\nl
&\quad
-{ ( 12 x_t^5 - 36 x_t^4 + 36 x_t^3 - 12 x_t^2 + 3 x_t - 2)\over 9 (x_t-1)^3}\log{x_t\over 1+x_t}
\nl
&\quad
- {2 x_t ( -3 + 7 x_t^2) \over 9(x_t^2-1)^3} \log 2- { 48 x_t^6 + 24 x_t^5 - 104 x_t^4 - 35 x_t^3 + 20 x_t^2 + 13 x_t + 18 \over 36(x_t^2-1)^2 (1+x_t)}\bigg]
  \nl
  &\quad
  + f_W\frac{m_W}{2 \pi M}\bigg[
  N_\ell\left( \frac83 \log{\mu\over m_W} - \frac13 \right)
 + \frac{16 x_t^4}{3(x_t^2-1)^3} \log{x_t} \log{\mu\over m_W}  - \frac{4(3x_t^2-1)}{3(x_t^2-1)^2}     \log{\mu \over m_W}
 \nl
 &\quad
 + \frac{16 x_t^2}{3} \log^2{x_t}
 - \frac{ 4( 4x_t^6-16x_t^4+6x_t^2+1)}{3(x_t^2-1)^3} \log{x_t}  + \frac{ 8x_t^2(x_t^6-3x_t^4+4x_t^2-1)}{3(x_t^2-1)^3} {\mathrm {Li}}_2(1-x_t^2)
 \nl
 &\quad
 + \frac{4\pi^2 x_t^2}{9}
 - \frac{8 x_t^4 - 7 x_t^2 + 1}{3(x_t^2-1)^2} 
 \bigg]+ \frac{f_Z}{16c_W}\bigg[\Big[2(c_V^{(U)2}+c_A^{(U)2})+3(c_V^{(D)2}+c_A^{(D)2})\Big]
\Big[-\frac{32}{9} \log{{\mu \over m_Z}} -4 \Big]\nl
&\quad+(c_V^{(U)2}+c_A^{(U)2})
\Big[\frac{32(24y_t^8-21y_t^6-4y_t^4+5y_t^2-1)}{9(4y_t^2-1)^{7/2}}\arctan(\sqrt{4y_t^2-1})-\frac{\pi y_t}{3}
\nl
&\quad +\frac{4(48 y_t^6+62 y_t^4-47 y_t^2+9)}{9(4y_t^2-1)^3}\Big]
+(c_V^{(U)2}-c_A^{(U)2})\Big[\frac{4y_t^2(624y_t^4-538y_t^2+103)}{9(4y_t^2-1)^3}-\frac{13\pi y_t}{3}
\nl
&\quad+\frac{32y_t^2(104y_t^6-91y_t^4+35y_t^2-5)}{3(4y_t^2-1)^{7/2}}\arctan(\sqrt{4y_t^2-1})\Big] \bigg] \nl
&\quad 
+\frac{f_Z}{24 c_W^2} \frac{m_W}{\pi M}
\bigg[\Big[2(c_V^{(U)2}+c_A^{(U)2})+3(c_V^{(D)2}+c_A^{(D)2})\Big]\Big(8\log{{\mu \over m_Z}}-1\Big)
\nl
&\quad -(c_V^{(U)2}+c_A^{(U)2})\Big[ 
\frac{1-18y_t^2+36y_t^4}{(4y_t^2-1)^2}
+\frac{8(1-4y_t^2+3y_t^4+18y_t^6)\log{y_t}}{(4y_t^2-1)^3}
 \nl
&\quad +\frac{16 y_t^2(2-13y_t^2+32 y_t^4 -18y_t^6)}{(4y_t^2-1)^{7/2}}
\big[2 \arctan{\Big( \frac{1}{\sqrt{4y_t^2-1}} \Big)} \log{y_t} 
- {\mathrm {Im}}\,{{\mathrm {Li}}_2\Big(\frac{1-i\sqrt{4y_t^2-1}}{2 y_t^2}\Big)} \big] \Big]
\nl
&\quad +4 y_t^2 (c_V^{(U)2}-c_A^{(U)2}) \Big[-\frac{8-59y_t^2+108 y_t^4}{(4y_t^2-1)^3} 
-\frac{(29-128y_t^2+108y_t^4)\log{y_t}}{(4y_t^2-1)^3} \nn\\
&\quad+ \frac{2(-7+38 y_t^2-82 y_t^4+108 y_t^6)}{(4y_t^2-1)^{7/2}}
\big[2 \arctan{\Big( \frac{1}{\sqrt{4y_t^2-1}} \Big)} \log{y_t} 
-{\mathrm {Im}}\,{{\mathrm {Li}}_2\Big(\frac{1-i\sqrt{4y_t^2-1}}{2 y_t^2}\Big)} \big]\Big]\bigg]
\Bigg\}
 \,,
\end{align}
where the reduced coefficients $\hat{c}_i^{(S)}$ are given in terms of the original Wilson coefficients by $c_i^{(S)}\equiv (\pi\alpha_2^2/m_W^3) \hat{c}^{(S)}_i$ with $\alpha_2=g_2^2/(4\pi)$,
where $i = u,d,s,c,b,g$ is the index for quark or gluon and $U$ denotes up-type while $D$ denotes down-type 
and we have neglected small corrections from $|V_{td}|^2$ and $|V_{ts}|^2$,
The $u$ and $c$ quarks have the same coefficients, as do $d$ and $s$ quarks through all the weak matching calculations.
The group theory factors are $f_W=J(J+1)-Y^2$, $f_Z=Y^2$.
The strong coupling is denoted by $\alpha_s(\mu)$. The mass ratios are defined as $x_j \equiv m_j/m_W$ and $y_j \equiv m_j/m_Z$ where $m_Z$ is the mass of $Z^0$ boson, and $j$ is the index of the specific particle, e.g. $j=t$ stands for top quark, $j=h$ for Higgs boson.
$\mathrm { Li}_{2}(z) \equiv \sum_{k = 1}^{\infty} {z^k}/{k^2}$ is the 
dilogarithm function.
$N_\ell=2$ is the number of massless Standard Model generations.

\section{Illustrative UV Completions \label{sec:uv_compeletion}}

At subleading order $1/M$, the underlying UV completion impacts spin-independent direct detection cross sections via the single parameter $c_H$, cf. Eq.\,(\ref{eq:results}).  This parameter is in turn determined by 
coefficients $A$, $B$, $C$, $A^\prime$, $B^\prime$ and $C^\prime$ in Eqs.\,(\ref{3ptvtx_even_pos}), (\ref{3ptvtx_even_neg}), and (\ref{3ptvtx_odd})
We illustrate the determination of $c_H$ by considering minimal Standard Model extensions with a new electroweak multiplet containing our dark matter WIMP. 

\subsection{Real bosons}

\begin{figure}[t]
        \captionsetup{width=.9\linewidth}
        \centering
        \includegraphics[width=.85\linewidth]{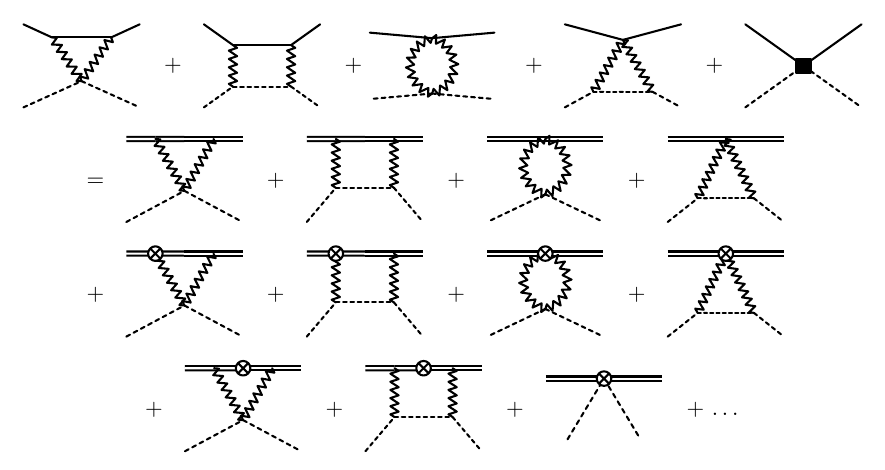}
\caption{\label{fig:chmatch_boson}
\small
  Matching condition for the coefficients in the EFT for UV theory consisting of
  the Standard Model plus an electroweak multiplet real boson.
  Solid lines denote the boson, dashed lines denote Higgs doublet,
  zigzag lines denote $\mathrm{SU(2)}_W\times \mathrm{U(1)}_Y$ gauge fields.  
  The black square denotes a counter term contact interaction.
  Matching is performed in the
  electroweak symmetric theory.  
  Double lines on the R.H.S. denote heavy WIMPs and the encircled cross denotes
  insertion of a $1/M$ effective theory vertex.
  }
\end{figure}

For a real boson electroweak multiplet, the matching between the UV theory and the effective theory to determine WIMP-Higgs interaction operators is shown in Fig.\,\ref{fig:chmatch_boson}. 
For a real scalar, we take the Lagrangian
\be
\mathcal{L}^{\mathrm{spin-0}}_{\mathrm{UV}}=\frac{1}{2}D_\mu \Phi D^\mu \Phi-\frac{1}{2}M^2\Phi^2\,,
\ee
and for a real vector, we take the Lagrangian
\be
\mathcal{L}^{\mathrm{spin-1}}_{\mathrm{UV}}=-\frac{1}{4}\left(D_\mu \mathcal{V}_\nu- D_\nu \mathcal{V}_\mu \right )\left(D^\mu \mathcal{V}^\nu- D^\nu \mathcal{V}^\mu \right)
+\frac{1}{2}M^2\mathcal{V}_\mu \mathcal{V}^\mu\,.
\ee

For a real scalar, a generalization of the results in Ref.~\cite{Chen:2018uqz} yields
\be
c_H^{\mathrm{spin-0}}=\frac{J(J+1)}{2}\alpha_2^2\log{\frac{\Lambda_{\rm UV}^2}{M^2}} + \dots \,,
\ee
where $\Lambda_{\rm UV}$ is a scale intrinsic to the UV theory 
(the ``UV theory of the UV theory" scale) and the ellipsis denotes terms that are not logarithmically enhanced in the limit $\Lambda_{\rm UV} \gg M$. 
In models such as a ``weakly interacting stable pion"~\cite{Bai:2010qg}, this scale is 
$\Lambda_{\rm UV}^2\sim M^2/\alpha_2$, and we consider this case in Sec.~\ref{sec:cs}. 
Similarly, for a real vector we find 
\be
c_H^{\mathrm{spin-1}}=\frac{J(J+1)}{6}\alpha_2^2\log{\frac{\Lambda_{\rm UV}^2}{M^2}}\,.
\ee

\subsection{Fermions}

For fermionic electroweak multiplets with hypercharge $Y=\pm1/2$, 
we consider a UV theory containing a Dirac field $\psi$, related
to the self-conjugate field $\chi$ as in Eq.~(\ref{eq:chidef}). 
We include another electroweak multiplet Majorana fermion $\chi^\prime$, 
in a $(J^\prime,\, Y^\prime)$ representation of 
$\mathrm{SU}(2)_W\times \mathrm{U}(1)_Y$, with 
mass $M^\prime \gg M$. 
The field $\chi^\prime$ serves to split the degenerate mass of the neutral components in $\chi$ 
(for $Y^\prime \ne 0$, the Majorana fermion $\chi^\prime$ is 
a reducible representation of $\mathrm{SU}(2)_W\times \mathrm{U}(1)_Y$ constructed from a Dirac fermion $\lambda$, 
similar to the construction of $\chi$ from $\psi$). 
We include the general renormalizable interaction $F(\psi,\,\chi^\prime,\, H)$  allowed by gauge invariance, 
\begin{align}
\mathcal{L}_{\mathrm { UV}}^{\mathrm{spin-1/2}}&=
\bar{\psi}(i\slashed{D}-M)\psi
+\frac{1}{2}\bar{\chi}^\prime (i\slashed{D}-M^\prime)\chi^\prime+F(\psi,\,\chi^\prime,\, H)\,,
\end{align}
where the detailed expression of $F(\psi,\,\chi^\prime,\, H)$ can be found in Appendix~\ref{sec:UVdetails}.

\begin{figure}[t]
\centering
        \includegraphics[width=.8\linewidth]{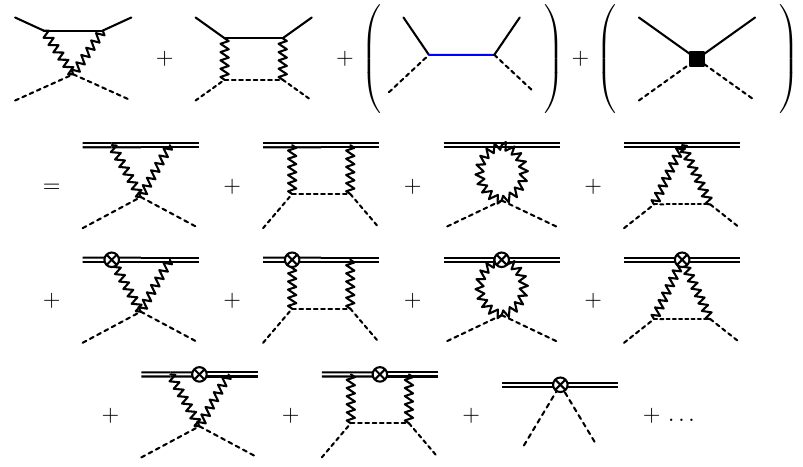}
          \captionsetup{width=.9\linewidth}
\caption{\label{fig:chmatch}
\small
  Matching condition for the coefficients in the EFT for UV theory consisting of
  the Standard Model plus electroweak multiplet self-conjugate fermion.
  Solid lines denote the fermions, with the blue line denoting the heavier fermion with mass $M^\prime$.
  Double lines denote the WIMPs.
  Other notations are the same as in the bosonic matching, Fig.\,\ref{fig:chmatch_boson}.
  The third diagram on the L.H.S only appears when $Y=\pm1/2$ for even-dimension electroweak multiplet.
  The last diagram on the L.H.S appears when the fermion is spin-3/2.
  }
\end{figure}

The matching is shown in Fig.\,\ref{fig:chmatch}.
The Higgs-WIMP-WIMP coupling is determined by the quantities $A$, $B$, $C$ in Eq.\,(\ref{ABCexprs}), which are given by explicit  computation as in Appendix~\ref{sec:UVdetails}.
In the pure WIMP limit, i.e. $M^\prime\gg M$, the Wilson coefficient in the Feynman rules Eq.\,(\ref{3ptvtx_even_pos}) and Eq.\,(\ref{3ptvtx_even_neg}) reduce to
\begin{align}
c_H=-\frac{3}{2}\alpha_2^2\left[J(J+1)+\tan^2{\theta_W} |Y|+\tan^4{\theta_W} Y^2\right]\,,
\label{evencH}
\end{align}
where $\theta_W$ is the weak mixing angle, and $Y=\pm 1/2$ for even-dimension electroweak multiplets.

For the  $Y=0$ case, we consider the limit $M^\prime \to \infty$
and the renormalizable Lagrangian containing a single electrically neutral component reduces to
\begin{align}
\mathcal{L}_{\mathrm { UV}}^{\mathrm{spin-1/2}}&=
\frac12 \bar{\chi}(i\slashed{D}-M)\chi\,.
\end{align}
The matching is again described by the diagrams in Fig.\,\ref{fig:chmatch}.
In the pure WIMP limit, we have $A^\prime=\frac{3}{2(4\pi)^2} g_2^4J(J+1)$ and $B^\prime=0$, and the Wilson coefficient from Eq.\,(\ref{3ptvtx_odd}) reduces to 
\begin{align}
c_H=-\frac{3}{2}\alpha_2^2 J(J+1)\,.
\end{align}

For a Rarita-Schwinger, spin-3/2, WIMP, we take the effective 
UV Lagrangian to be
\be
\mathcal{L}_{\mathrm{UV}}^{\mathrm{spin-3/2}}=-\bar{\varPsi}^\mu\left[\left(i\slashed{D}-M\right)g_{\mu\nu}
-\left(i\gamma_\mu D_\nu+i \gamma_\nu D_\mu\right)
+\gamma_\mu\left(i\slashed{D}+M\right)\gamma_\nu
\right]{\varPsi}^\nu\,,
\ee
and do the matching as in Fig.\,\ref{fig:chmatch}, similar to the procedures for the spin-1/2 case.
We obtain the coefficient
\ba
c_H=-\frac{2}{3}\alpha_2^2\left[J(J+1)+\tan^2{\theta_W} |Y|+\tan^4{\theta_W} Y^2\right]\log{\frac{\Lambda_{\rm UV}^2}{M^2}} + \dots \,,
\ea
where $Y=0$ for odd-dimension multiplets and $Y=\pm 1/2$ for even-dimension multiplets.

\section{Cross Sections \label{sec:cs}}

The benchmark WIMP-nucleon elastic scattering spin-independent cross section is
\begin{align}
\sigma_N 
=\frac{m_{r}^{2}}{\pi}|\mathcal{M}^{(0)}_{N} + \mathcal{M}^{(2)}_{N}|^{2}\,,
\end{align}
where $N=n,p$ is a nucleon, $m_r = m_N M /(m_N + M) \approx m_N$ is the reduced mass of
the WIMP-nucleon system, and the scattering amplitude is
\begin{equation}
\mathcal{M}^{(S)}_{N} = \sum_{i= q,g} c^{(S)}_{i}(\mu_{0}) \langle N | O^{(S)}_{i}(\mu_{0})|N\rangle \,,
\label{amp3f}
\end{equation}
where $S=0, 2$ for operators with different spins. 
The nucleon states $|N\rangle$ are non-perturbative and we use Lattice QCD to evaluate the nucleon matrix elements at energy scale $\mu_0 \sim$ GeV.
So the heavy quarks, bottom and charm need to be integrated out from the 5-flavor QCD theories, Eq.\,(\ref{5fQCD}).
Renormalization group evolution from the 5-flavor effective QCD theory at the weak scale $\mu_{t}$  to the bottom quark mass scale $\mu_b \sim m_b$,
threshold matching at $\mu_b$, running from $\mu_b$ to charm quark mass scale $\mu_c \sim m_c$, 
threshold matching at $\mu_c$, further running from $\mu_c$ to $\mu_0 $ are performed.  Details can be found in Ref.~\cite{Hill:2014yxa}.
Specifically, we take 
$\mu_{t} = (m_t+m_W)/2 = 126\,{\rm GeV}$, 
$\mu_{b} = 4.75\,{\rm GeV}$, 
$\mu_{c} = 1.4\,{\rm GeV}$, and 
$\mu_{0} = 1.2\,{\rm GeV}$.
For the spin-0 coefficients, renormalization group evolution and threshold matching are performed at NNNLO.
For spin-2 coefficients, the running and matching are at NLO.
In the end, we obtain the 3-flavor effective QCD theory,
with $q=u, d, s$ in Eq.\,(\ref{amp3f}) being the three light flavors and $g$ denoting the gluon.
We take the same Lattice QCD data for nucleon matrix elements as in Ref.~\cite{Chen:2019gtm}.

\begin{figure}[htb]
\centering
        \captionsetup{width=0.9\linewidth}
\includegraphics[width=.45\linewidth]{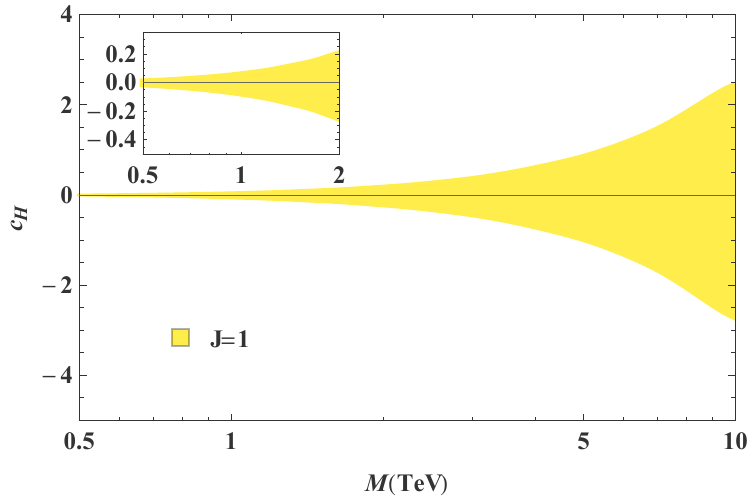}
\includegraphics[width=.45\linewidth]{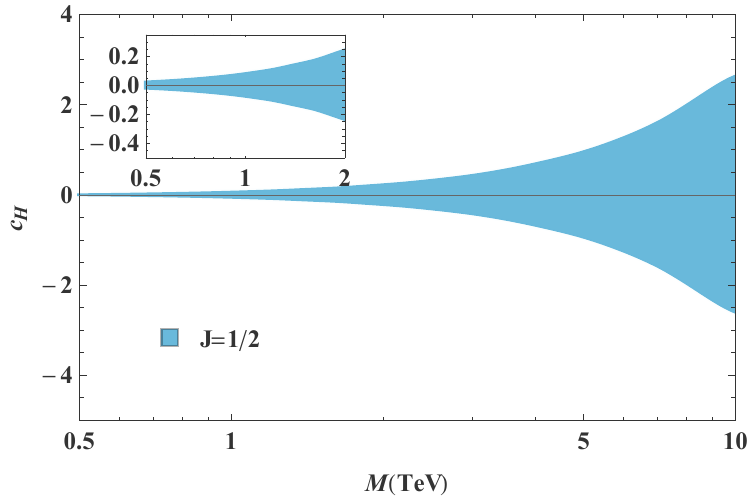}
\includegraphics[width=.45\linewidth]{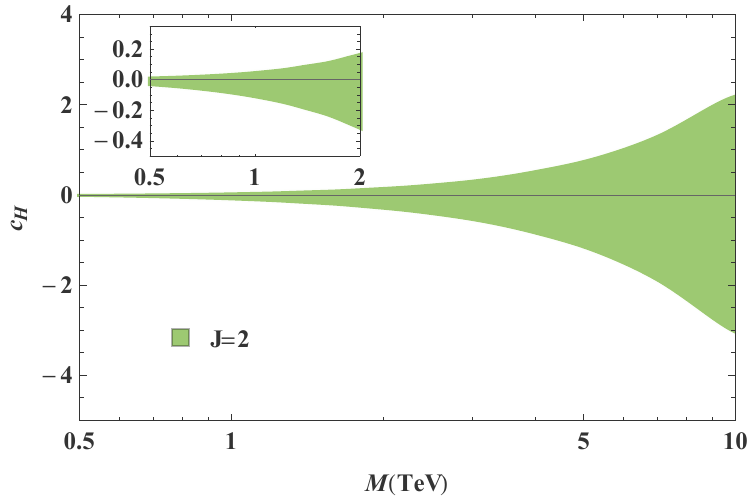}
\includegraphics[width=.45\linewidth]{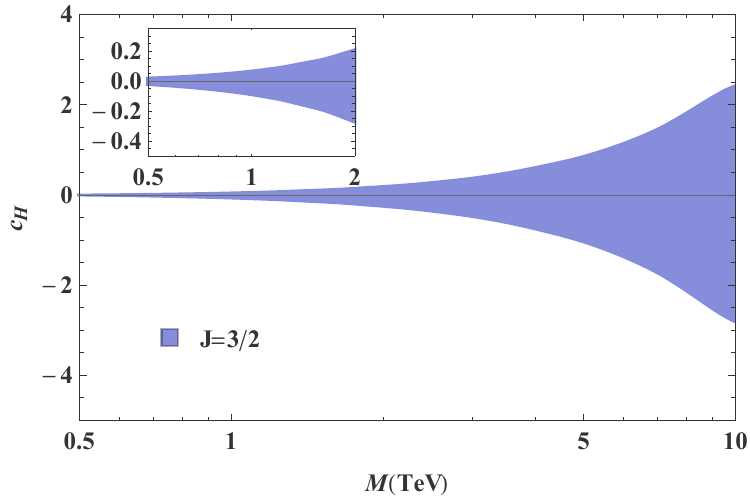}
\includegraphics[width=.45\linewidth]{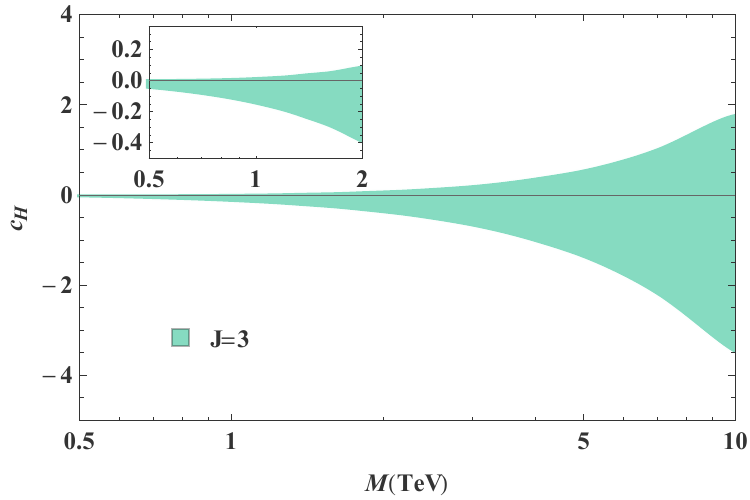}
\includegraphics[width=.45\linewidth]{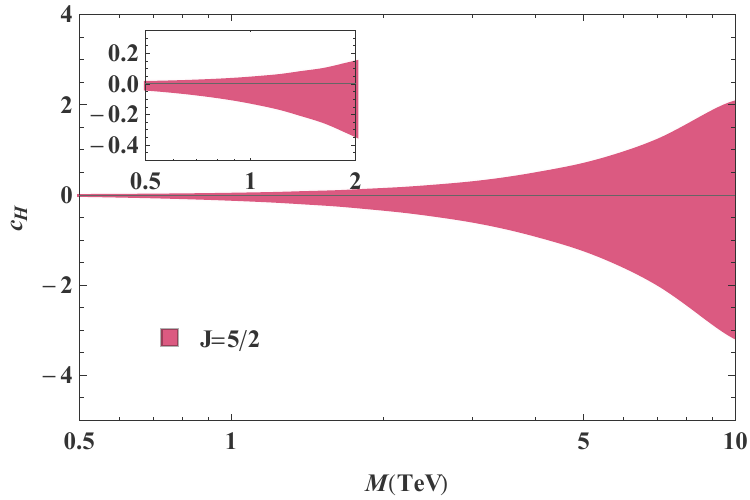}
\caption{\small
Constraints on the dimensionless parameter $c_H$ for WIMPS with different isospins, versus WIMP mass.
Zoomed bounds for WIMP mass smaller than 2 TeV are exhibited at the upper left corners.
}
\label{purecH_lows}
\end{figure}

For our default matching scales $\mu_t$, $\mu_b$, $\mu_c$ and $\mu_0$, and with the central values of all nucleon matrix elements at scale $\mu_0$, we find that the spin-0 and spin-2 amplitudes for WIMP and proton scattering are (normalized by 
spin-2 amplitude  ${\mathcal M}^{(2)}_p|_{M\to\infty} = J(J+1)$ when $Y=0$)
\begin{align}
  {\mathcal M}_p^{(0)} &= -0.82J(J+1) -0.42Y^2-299.50 c_H{m_W \over M} \,,
  \nl
  {\mathcal M}_p^{(2)} &=  J(J+1)-0.25Y^2 - \left[0.51J(J+1)-0.072 Y^2\right]  {m_W\over M} \,,
        \label{genM2}
\end{align}
where the low energy effective theory of WIMP and 3-flavor QCD operators at $1/M$ order is yet to be determined by one parameter $c_H$.
We may constrain $c_H$ by current direct detection experimental limits~\cite{LZ:2022lsv}.
We plot the allowed region of $c_H$ for different isospins of a WIMP in Fig.\,\ref{purecH_lows}.

We may match onto the minimal UV theories in Sec.~\ref{sec:uv_compeletion} to obtain concrete values for $c_H$ and predict benchmark results for general WIMP and nucleon spin-independent scattering cross sections.
For a real bosonic heavy WIMP, the central values for the amplitudes are
\begin{align}
  {\mathcal M}_p^{(0)} &= -J(J+1)\left(0.824-0.342\eta{m_W \over M}\log{\Lambda_{\rm UV} \over M}\right)\,,
   \nl
  {\mathcal M}_p^{(2)} &=  J(J+1)\left(1-0.515 {m_W\over M}\right)\,,
     \label{bM}
\end{align}
where $\eta=1$ for a spin-0 WIMP, $\eta=1/3$ for a spin-1 WIMP, and $\Lambda_{\rm UV}$ is a UV scale.  We take $\Lambda_{\rm UV}\sim M/\sqrt{\alpha_2} \sim 10\,M$ for illustration, as discussed in Sec.~\ref{sec:uv_compeletion}.
For a self-conjugate spin-1/2 heavy WIMP, the amplitudes are
\begin{align}
  {\mathcal M}_p^{(0)} &= -0.824J(J+1) -0.417Y^2+ \left[0.513J(J+1)+0.153|Y|+ 0.0457Y^2\right]{m_W \over M} \,,
  \nl
  {\mathcal M}_p^{(2)} &=  J(J+1)-0.247Y^2 - \left[0.515J(J+1)-0.0716 Y^2\right]  {m_W\over M} \,.
    \label{12M}
\end{align}
For a self-conjugate spin-3/2 heavy WIMP, the amplitudes are
\begin{align}
  {\mathcal M}_p^{(0)} &= -0.824J(J+1) -0.417Y^2+ \left[0.456J(J+1)+0.136|Y|+ 0.0407Y^2\right]{m_W \over M}\log{\Lambda_{\rm UV} \over M} \,,
  \nl
  {\mathcal M}_p^{(2)} &=  J(J+1)-0.247Y^2 - \left[0.515J(J+1)-0.0716 Y^2\right]  {m_W\over M} \,,
   \label{32M}
\end{align}
where again $\Lambda_{\rm UV}$ is a UV scale 
and we will take it to be $M/\sqrt{\alpha_2}$.
From expressions (\ref{bM}), (\ref{12M}) and (\ref{32M})
we see clearly the cancellation between spin-0 and spin-2 amplitudes.  
For all values of spin and for all electroweak quantum numbers with $J(J+1)\ge Y^2$ (such that the multiplet contains an electrically neutral component),
    ${\cal M}^{(0)}$ 
    is negative at leading power and ${\cal M}^{(2)}$ is positive. 
Similarly at $1/M$ order, the contributions to ${\cal M}^{(0)}$ and ${\cal M}^{(2)}$ have opposite sign.  The cancellation is especially severe for 
the Higgsino-like case $J=Y=1/2$.  

\begin{figure}[t]
\centering
        \captionsetup{width=0.9\linewidth}
\includegraphics[width=.49\linewidth]{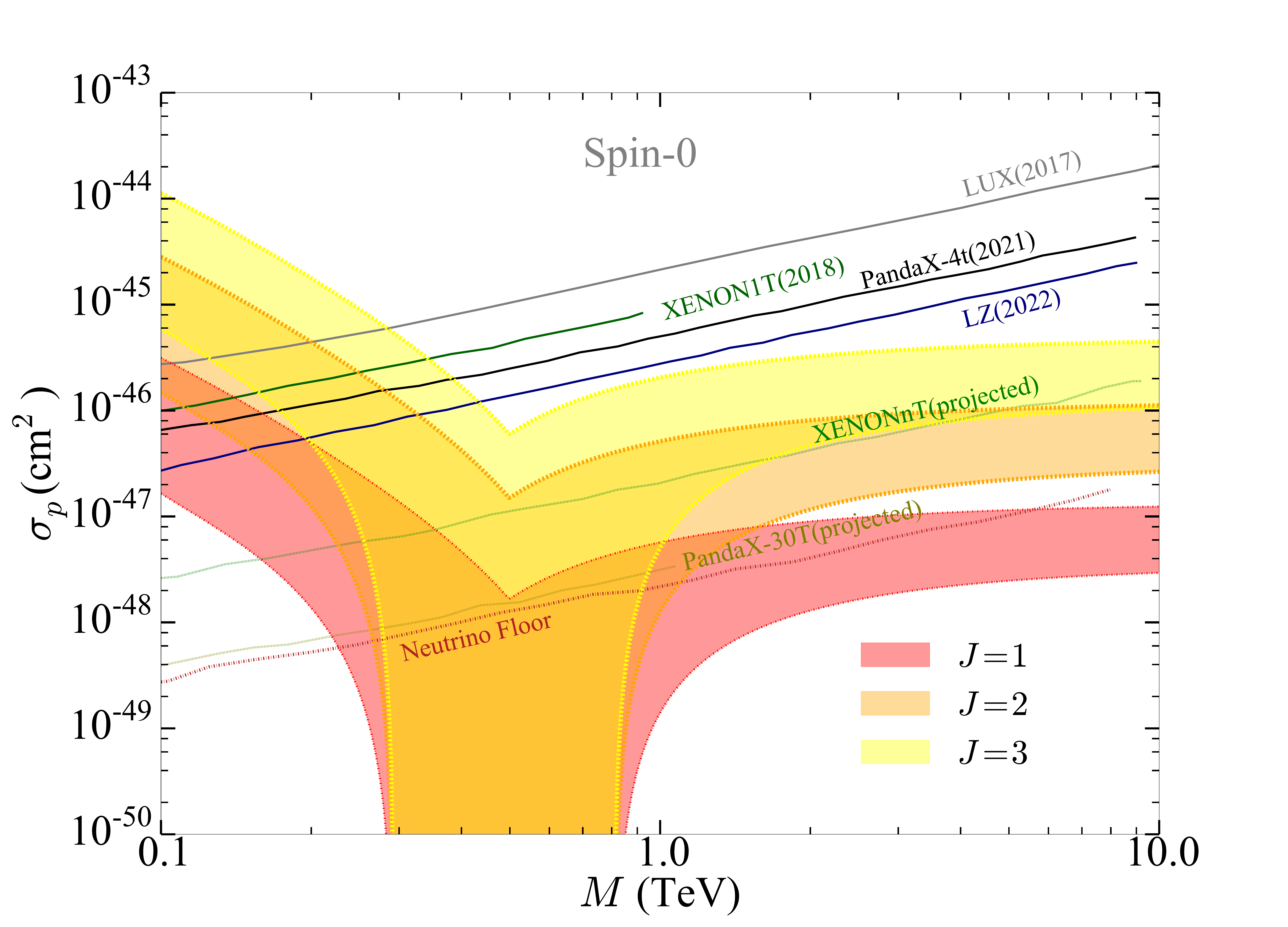}
\includegraphics[width=.49\linewidth]{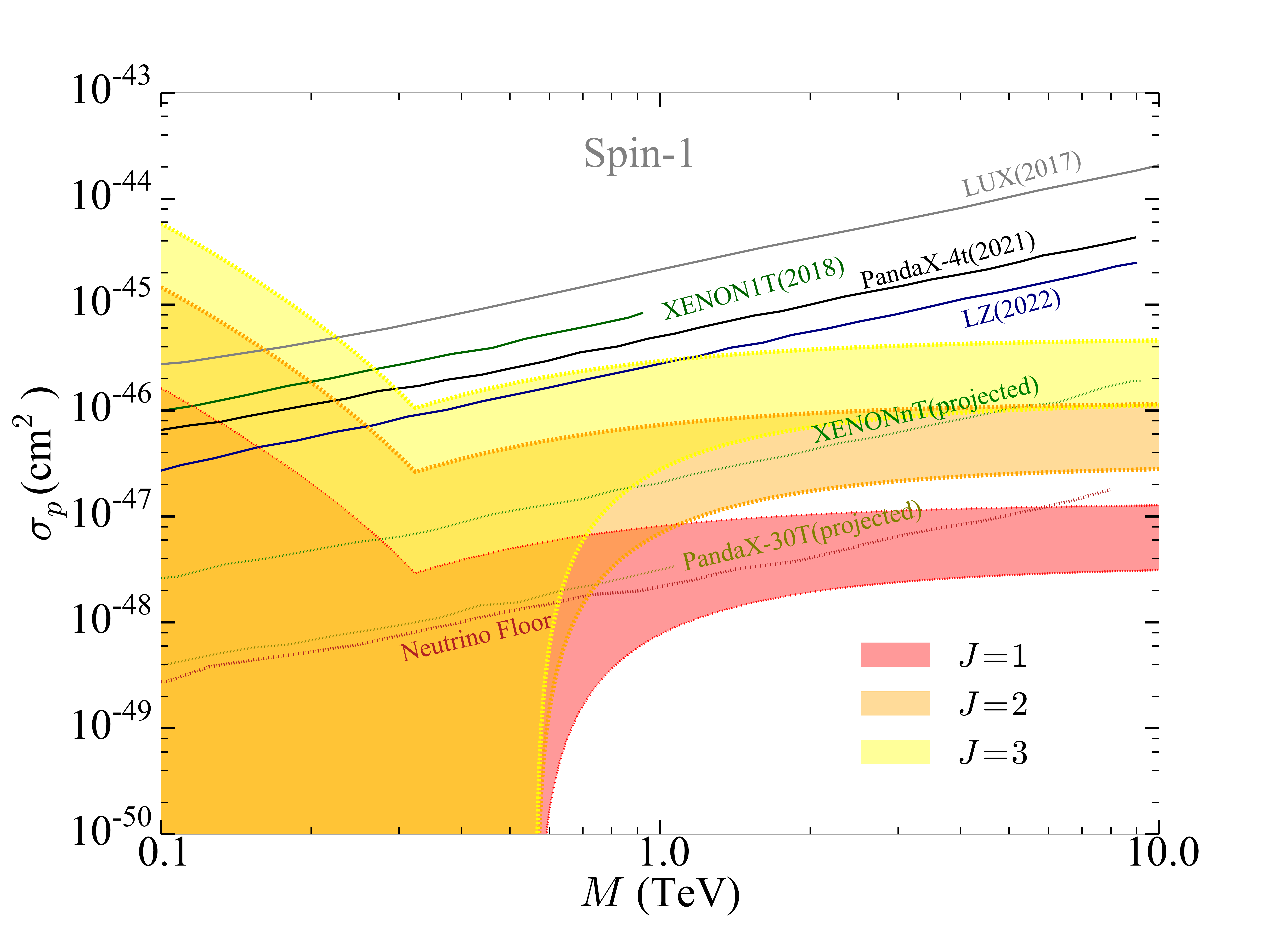}
\caption{\small
Spin-independent scattering cross section for different bosonic WIMP multiplets on proton, versus the WIMP mass.
}
\label{gen_cs_boson}
\end{figure}

\begin{figure}[t]
\centering
        \captionsetup{width=0.9\linewidth}
\includegraphics[width=.49\linewidth]{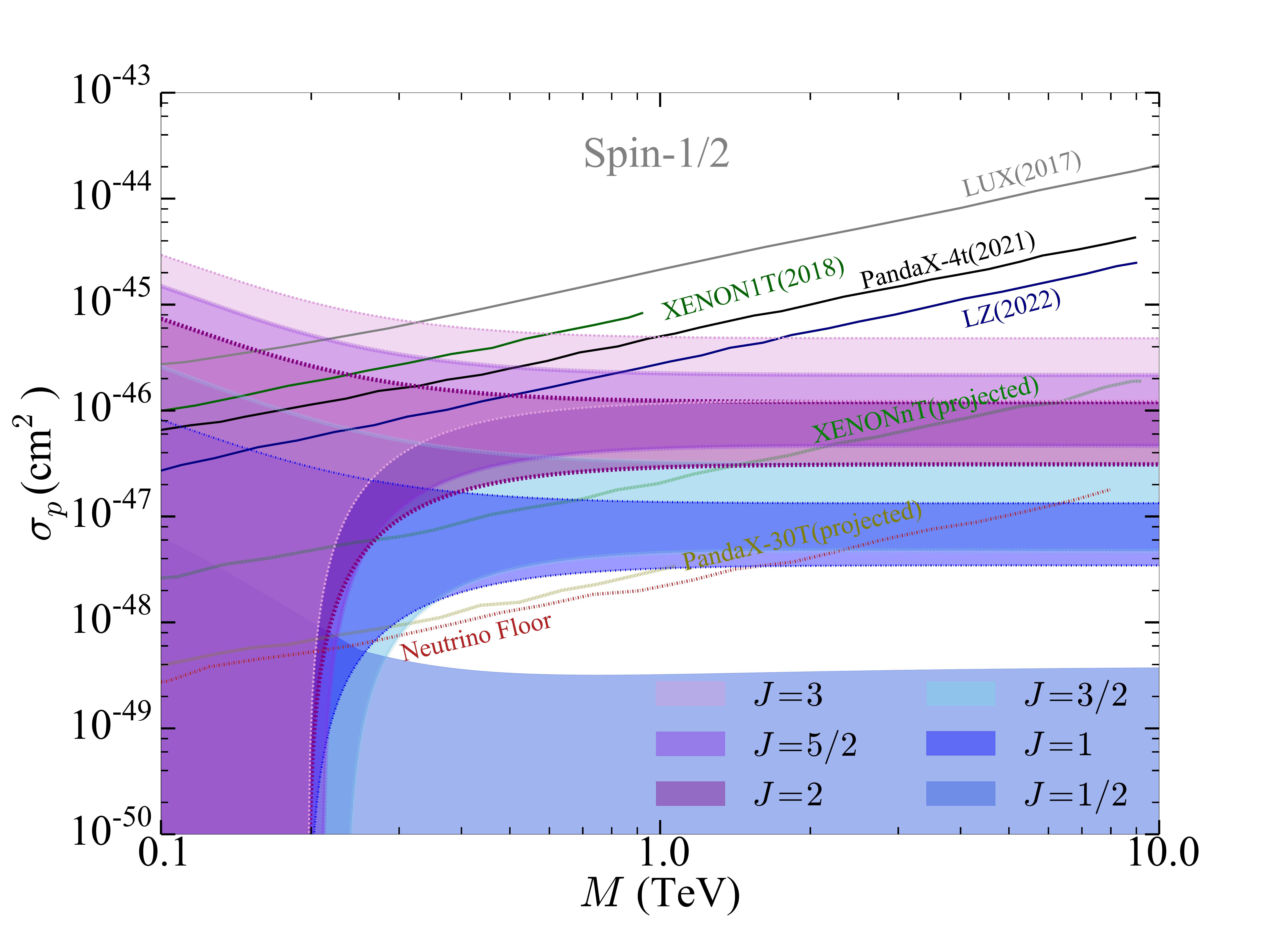}
\includegraphics[width=.49\linewidth]{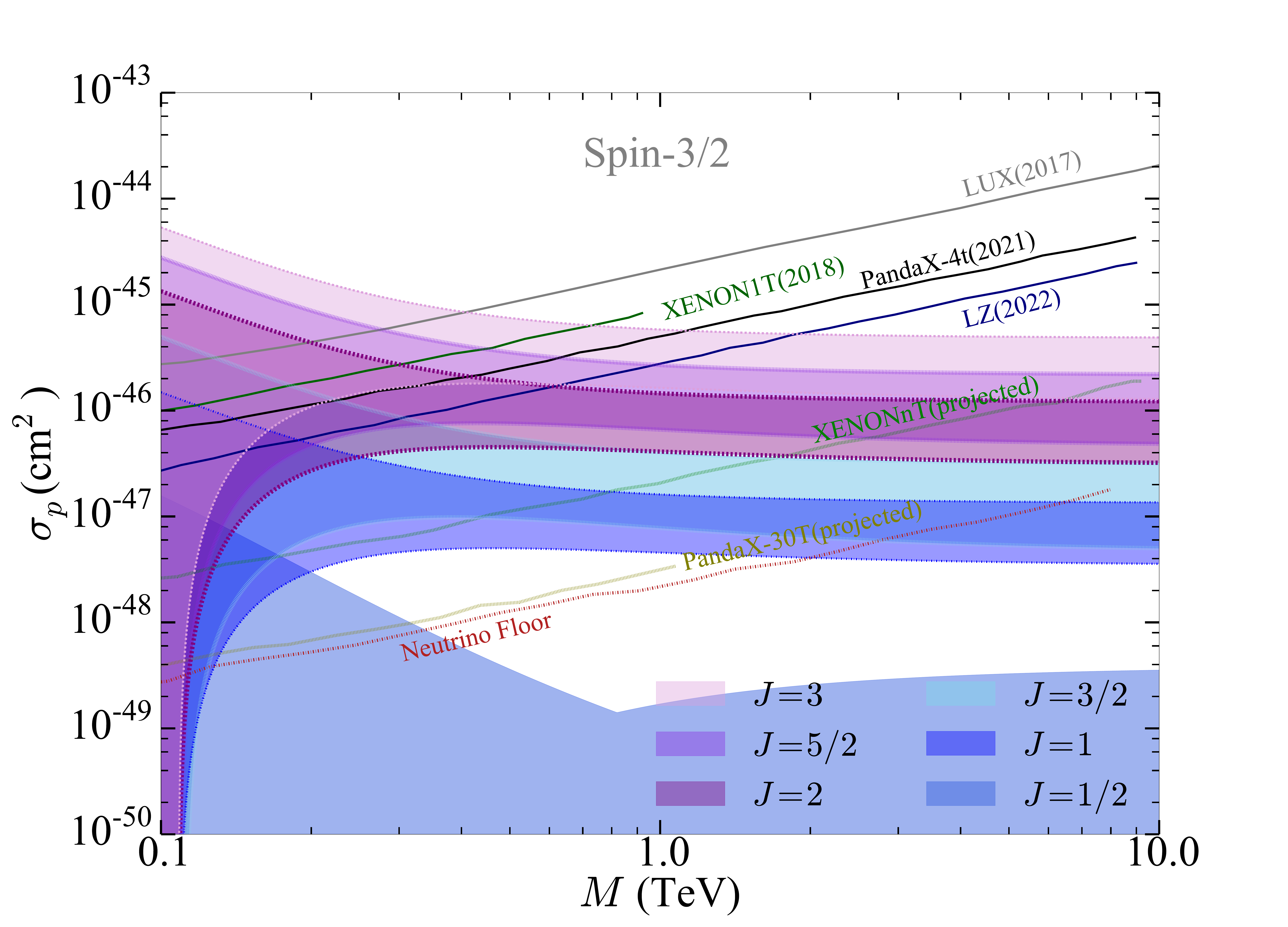}
\caption{\small
Spin-independent scattering cross section for different fermionic WIMP multiplets on proton, versus the WIMP mass.
}
\label{gen_cs_fermion}
\end{figure}

We plot the spin-independent cross sections for different heavy WIMPs and proton scattering in Fig.\,\ref{gen_cs_boson} and Fig.\,\ref{gen_cs_fermion}, 
versus the mass of the WIMP.
When evaluating the nucleon-level amplitude, we have perturbative uncertainties from Wilson coefficients and non-perturbative uncertainties from hadronic matrix elements.
Uncertainty for perturbative computation of the matching coefficients is estimated by 
varying the matching scales within the ranges $m_W^2/2 \le \mu_t^2 \le 2 m_t^2$,
$m_b^2/2 \le \mu_b^2 \le 2 m_b^2$,
$m_c^2/2 \le \mu_c^2 \le 2 m_c^2$,
and
$1.0\,{\rm GeV} \le \mu_0 \le 1.4\,{\rm GeV}$. 
Uncertainties from neglect of higher order (starting from $1/M^2$ order) power corrections are estimated
by shifting ${\cal M}^{(2)}_p \to {\cal M}^{(2)}_p|_{M\to\infty}[ 1 \pm (m_W/M)^2 ]$.
Uncertainties from nucleon matrix elements are propagated to the observable cross section~\cite{Gasser:1982ap, Durr:2011mp, Junnarkar:2013ac, Martin:2009iq, Hill:2014yxa}. 
We add the errors in quadrature from different sources mentioned above, for spin-0 and spin-2 amplitudes separately.  
 The maximum and minimum of all possible values of the combination $|\mathcal{M}^{(0)}_{p} + \mathcal{M}^{(2)}_{p}|$ set the bounds of the colored cross section bands for each WIMP in Fig.\,\ref{gen_cs_boson} and Fig.\,\ref{gen_cs_fermion}.
 The cross section increases as the isospin increases, and the central value varies from $10^{-50}\, \rm{cm}^2$ order to $10^{-46}\, \rm{cm}^2$ order
from lowest isospin $J=1/2$ to highest isospin $J=3$ for WIMP mass at 1 TeV.

In Fig.\,\ref{gen_cs_boson} and Fig.\,\ref{gen_cs_fermion}, also shown are the recent dark matter direct detection experimental 
exclusion ($90\%$ confidence) limits \cite{LZ:2022lsv, PandaX-4T:2021bab, Aprile:2018dbl, Akerib:2016vxi} for the relevant TeV mass range.
Above the TeV scale, for isospin $J$ smaller than 3, the WIMP proton cross section is below current experimental sensitivity.
For fermionic WIMPs, when $J=3$, the cross section overlaps with LZ's limit in the 1~TeV mass region.
For bosonic triplet ($J=1$) WIMPs, the cross section is close to the neutrino floor~\cite{Billard:2013qya}.
For fermionic doublet ($J=1/2$) WIMPs, the cross section upper bound is much lower than the neutrino floor.
Other low-isospin WIMPs lie between current experimental limit and the neutrino floor.

\section{Summary \label{sec:summ}}
We have used heavy particle effective theory to study general heavy WIMP and nucleon scattering at subleading $1/M$ order, and to
compute cross sections for arbitrary electroweak representations and low-spin particles.
We focused on the elastic and spin-independent process which is a primary target for dark matter direct detection experiments.  

The spin-independent cross section is universal at leading power, determined by Standard Model parameters once the WIMP spin and electroweak representation are specified.  
At subleading $1/M$ order, dependence on UV structure is encoded by
a single coefficient $c_H$ describing the WIMP-WIMP-Higgs boson coupling. 
We constrain this coefficient using current experimental exclusion limits~\cite{LZ:2022lsv} 
and find $-0.2 \lesssim c_H \lesssim 0.1$ (90\% CL) 
at 1 TeV WIMP mass, with a slight 
dependence on the isospin of the WIMP, cf. Fig.~\ref{purecH_lows}.
These model-independent results can be interpreted as constraints 
on the parameter space of specific UV completions, cf. e.g. Ref.~\cite{Ellis:2023ndh}. 

We also predict benchmark cross sections in dark matter direct detection experiments through $1/M$ order, by matching our heavy WIMP effective theory to minimal UV extensions of Standard Model 
to obtain $c_H$.
The corresponding cross sections are below current experimental limits for low isospin electroweak multiplets, either bosonic or fermionic, 
mostly lying between the experimental limit and the neutrino floor.  Central values vary between $\sim 10^{-50}\, \rm{cm}^2$ and $\sim 10^{-46}\, \rm{cm}^2$
from lowest isospin $J=1/2$ to highest isospin $J=3$ for WIMP mass at 1 TeV.  
These cross sections are within 
striking range of next-generation experiments, with the exception of electroweak doublets, 
which hide below the neutrino floor. 
In general, higher-isospin WIMPs have larger cross section and 
will be discovered or excluded first. 
Uncertainties due to nuclear effects of the heavy element experimental target, e.g. Xenon, are similar in magnitude to  uncertainties of the cross section which have been computed here, and are not expected to change the predicted discovery range for these WIMPs~\cite{Chen:2019gtm}. 
For most cases these heavy WIMPs can be discovered or excluded 
with next-generation direct detection experiments. 
An exceptional case is the electroweak doublet, whose cross section is impacted by a severe amplitude cancellation.  
Experimental methods such as directional discrimination, annual modulation and improved flux measurements~\cite{Billard:2009mf, OHare:2020lva, Vahsen:2021gnb} may allow access to cross sections below the neutrino floor, complementing indirect searches~\cite{Co:2021ion,Dessert:2022evk}. 

\vskip 0.2in
\noindent
{\bf Acknowledgements.} 
QC acknowledges postdoctorate fellowship supported by University of Science and Technology of China and Peng Huanwu Center for Fundamental Theory (PCFT), Hefei.
PCFT is supported by National Natural Science Foundation of China under grant No.\,12247103.
GJD is supported by the National Natural Science Foundation of China under Grant
Nos. 11975224, 11835013.
Research of RJH supported by the U.S. Department of Energy, Office of Science, Office of High Energy Physics, under Award Number DE-SC0019095.
This manuscript has been authored by Fermi Research Alliance, LLC under Contract No. DE-
AC02-07CH11359 with the U.S. Department of Energy, Office of Science, Office of High Energy Physics.

\vskip 0.1in
\noindent

\appendix

\section{Details of UV matching onto HWET for fermions \label{sec:UVdetails}}

Consider the Lagrangian,
\begin{align}
\mathcal{L}_{\mathrm { UV}}^{\mathrm{spin-1/2}}&=
\bar{\psi}(i\slashed{D}-M)\psi
+\frac{1}{2}\bar{\chi}^\prime (i\slashed{D}-M^\prime)\chi^\prime
\nl
+&\delta_{Y\prime,\,Y-1/2}\Bigg\{
\bar{\psi}_{k}\bigg[A_1\sqrt{k-1}\delta_{l,\,k-1}\delta_{J^\prime,\,J-1/2}
+B_1\sqrt{2J+2-k}\delta_{lk}\delta_{J^\prime,\,J+1/2}\bigg]\lambda_{l}H_2
\nl
&\quad-\bar{\psi}_{k}\bigg[-A_1\sqrt{2J+1-k}\delta_{l,\,k-1}\delta_{J^\prime,\,J-1/2}
+B_1\sqrt{k}\delta_{l,\,k+1}\delta_{J^\prime,\,J+1/2}
\bigg]\lambda_{l}H_1
\nl
&\quad+\bar{\psi}_{k}\bigg[A_1^\prime\sqrt{k-1}\delta_{l,\,k-1}\delta_{J^\prime,\,J-1/2}
+B_1^\prime\sqrt{2J+2-k}\delta_{lk}\delta_{J^\prime,\,J+1/2}\bigg]\gamma^5\lambda_{l}H_2
\nl
&\quad-\bar{\psi}_{k}\bigg[-A_1^\prime\sqrt{2J+1-k}\delta_{l,\,k-1}\delta_{J^\prime,\,J-1/2}
+B_1^\prime\sqrt{k}\delta_{l,\,k+1}\delta_{J^\prime,\,J+1/2}
\bigg]\gamma^5\lambda_{l}H_1
\nl
&\quad-\bar{\psi_{k}^{c}}\bigg[A_8\sqrt{2J+1-k}\delta_{kl}\delta_{J^\prime,\,J-1/2}
-B_8\sqrt{k}\delta_{l,\,k+1}\delta_{J^\prime,\,J+1/2}\bigg]\lambda^c_{l}H_1^*
\nl
&\quad-\bar{\psi_{k}^{c}}\bigg[A_8\sqrt{k-1}\delta_{l,\,k-1}\delta_{J^\prime,\,J-1/2}
+B_8\sqrt{2J+2-k}\delta_{lk}\delta_{J^\prime,\,J+1/2}
\bigg]\lambda^c_{l}H_2^*
\nl
&\quad-\bar{\psi_{k}^{c}}\bigg[A_8^\prime\sqrt{2J+1-k}\delta_{kl}\delta_{J^\prime,\,J-1/2}
-B_8^\prime\sqrt{k}\delta_{l,\,k+1}\delta_{J^\prime,\,J+1/2}\bigg]\gamma^5\lambda^c_{l}H_1^*
\nl
&\quad-\bar{\psi_{k}^{c}}\bigg[A_8^\prime\sqrt{k-1}\delta_{l,\,k-1}\delta_{J^\prime,\,J-1/2}
+B_8^\prime\sqrt{2J+2-k}\delta_{lk}\delta_{J^\prime,\,J+1/2}
\bigg]\gamma^5\lambda^c_{l}H_2^*
\Bigg\}
\nl
&\quad+\delta_{Y\prime,\,Y+1/2}\Bigg\{
-\bar{\psi}_k\bigg[A_2\sqrt{k-1}\delta_{l,\,k-1}\delta_{J^\prime,\,J-1/2}
+B_2\sqrt{2J+2-k}\delta_{lk}\delta_{J^\prime,\,J+1/2}\bigg]\lambda_{l}H_1^*
\nl
&\quad-\bar{\psi}_k\bigg[-A_2\sqrt{2J+1-k}\delta_{l,\,k-1}\delta_{J^\prime,\,J-1/2}
+B_2\sqrt{k}\delta_{l,\,k+1}\delta_{J^\prime,\,J+1/2}
\bigg]\lambda_{l}H_2^*
\nl
&\quad-\bar{\psi}_k\bigg[A_2^\prime\sqrt{k-1}\delta_{l,\,k-1}\delta_{J^\prime,\,J-1/2}
+B_2^\prime\sqrt{2J+2-k}\delta_{lk}\delta_{J^\prime,\,J+1/2}\bigg]\gamma^5\lambda_{l}H_1^*
\nl
&\quad-\bar{\psi}_k\bigg[-A_2^\prime\sqrt{2J+1-k}\delta_{l,\,k-1}\delta_{J^\prime,\,J-1/2}
+B_2^\prime\sqrt{k}\delta_{l,\,k+1}\delta_{J^\prime,\,J+1/2}
\bigg]\gamma^5\lambda_{l}H_2^*
\nl
&\quad+\bar{\psi_{k}^{c}}\bigg[A_6\sqrt{2J+1-k}\delta_{kl}\delta_{J^\prime,\,J-1/2}
-B_6\sqrt{k}\delta_{l,\,k+1}\delta_{J^\prime,\,J+1/2}\bigg]\lambda^c_{l}H_2
\nl
&\quad-\bar{\psi_{k}^{c}}\bigg[A_6\sqrt{k-1}\delta_{l,\,k-1}\delta_{J^\prime,\,J-1/2}
+B_6\sqrt{2J+2-k}\delta_{lk}\delta_{J^\prime,\,J+1/2}
\bigg]\lambda^c_{l}H_1
\nl
&\quad+\bar{\psi_{k}^{c}}\bigg[A_6^\prime\sqrt{2J+1-k}\delta_{kl}\delta_{J^\prime,\,J-1/2}
-B_6^\prime\sqrt{k}\delta_{l,\,k+1}\delta_{J^\prime,\,J+1/2}\bigg]\gamma^5\lambda^c_{l}H_2
\nl
&\quad-\bar{\psi_{k}^{c}}\bigg[A_6^\prime\sqrt{k-1}\delta_{l,\,k-1}\delta_{J^\prime,\,J-1/2}
+B_6^\prime\sqrt{2J+2-k}\delta_{lk}\delta_{J^\prime,\,J+1/2}
\bigg]\gamma^5\lambda^c_{l}H_1
\Bigg\}
\nl
&\quad+\delta_{Y\prime,\,-Y+1/2}\Bigg\{
\bar{\psi}_k (-1)^k\bigg[A_3\sqrt{k-1}\delta_{l+k,\,2J+2}\delta_{J^\prime,\,J-1/2}
\nl
&\quad -B_3\sqrt{2J+2-k}\delta_{k+l,\,2J+3}\delta_{J^\prime,\,J+1/2}\bigg]\lambda^c_{l}H_2
\nl
&\quad-\bar{\psi}_k (-1)^k \bigg[A_3\sqrt{2J+1-k}\delta_{l+k,\,2J+1}\delta_{J^\prime,\,J-1/2}
+B_3\sqrt{k}\delta_{l+k,\,2J+2}\delta_{J^\prime,\,J+1/2}
\bigg]\lambda^c_{l}H_1
\nl
&\quad +\bar{\psi}_k (-1)^k\bigg[A_3^\prime\sqrt{k-1}\delta_{l+k,\,2J+2}\delta_{J^\prime,\,J-1/2}
-B_3^\prime\sqrt{2J+2-k}\delta_{k+l,\,2J+3}\delta_{J^\prime,\,J+1/2}\bigg]\gamma^5\lambda^c_{l}H_2
\nl
&\quad-\bar{\psi}_k (-1)^k\bigg[A_3^\prime\sqrt{2J+1-k}\delta_{l+k,\,2J+1}\delta_{J^\prime,\,J-1/2}
+B_3^\prime\sqrt{k}\delta_{l+k,\,2J+2}\delta_{J^\prime,\,J+1/2}
\bigg]\gamma^5\lambda^c_{l}H_1
\nl
&\quad-\bar{\psi_{k}^{c}} (-1)^k\bigg[A_7\sqrt{2J+1-k}\delta_{k+l,\,2J+1}\delta_{J^\prime,\,J-1/2}
+B_7\sqrt{k}\delta_{k+l,\,2J+2}\delta_{J^\prime,\,J+1/2}
\bigg]\lambda_{l}H_1^*
\nl
&\quad-\bar{\psi_{k}^{c}} (-1)^k\bigg[-A_7\sqrt{k-1}\delta_{k+l,\,2J+2}\delta_{J^\prime,\,J-1/2}
+B_7\sqrt{2J+2-k}\delta_{k+l,\,2J+3}\delta_{J^\prime,\,J+1/2}
\bigg]\lambda_{l}H_2^*
\nl
&\quad-\bar{\psi_{k}^{c}} (-1)^k\bigg[A_7^\prime\sqrt{2J+1-k}\delta_{k+l,\,2J+1}\delta_{J^\prime,\,J-1/2}
+B_7^\prime\sqrt{k}\delta_{k+l,\,2J+2}\delta_{J^\prime,\,J+1/2}
\bigg]\gamma^5\lambda_{l}H_1^*
\nl
&\quad-\bar{\psi_{k}^{c}} (-1)^k\bigg[-A_7^\prime\sqrt{k-1}\delta_{k+l,\,2J+2}\delta_{J^\prime,\,J-1/2}
+B_7^\prime\sqrt{2J+2-k}\delta_{k+l,\,2J+3}\delta_{J^\prime,\,J+1/2}
\bigg]\gamma^5\lambda_{l}H_2^*
\Bigg\}
\nl
&\quad+\delta_{Y\prime,\,-Y-1/2}\Bigg\{
-\bar{\psi}_k (-1)^k\bigg[A_4\sqrt{k-1}\delta_{l+k,\,2J+2}\delta_{J^\prime,\,J-1/2}
\nl
&\quad-B_4\sqrt{2J+2-k}\delta_{k+l,\,2J+3}\delta_{J^\prime,\,J+1/2}\bigg]\lambda^c_{l}H_1^*
\nl
&\quad-\bar{\psi}_k (-1)^k\bigg[-A_4\sqrt{2J+1-k}\delta_{l+k,\,2J+1}\delta_{J^\prime,\,J-1/2}
+B_4\sqrt{k}\delta_{l+k,\,2J+2}\delta_{J^\prime,\,J+1/2}
\bigg]\lambda^c_{l}H_2^*
\nl
&\quad -\bar{\psi}_k (-1)^k\bigg[A_4^\prime\sqrt{k-1}\delta_{l+k,\,2J+2}\delta_{J^\prime,\,J-1/2}
-B_4^\prime\sqrt{2J+2-k}\delta_{k+l,\,2J+3}\delta_{J^\prime,\,J+1/2}\bigg]\gamma^5\lambda^c_{l}H_1^*
\nl
&\quad-\bar{\psi}_k (-1)^k\bigg[-A_4^\prime\sqrt{2J+1-k}\delta_{l+k,\,2J+1}\delta_{J^\prime,\,J-1/2}
+B_4^\prime\sqrt{k}\delta_{l+k,\,2J+2}\delta_{J^\prime,\,J+1/2}
\bigg]\gamma^5\lambda^c_{l}H_2^*
\nl
&\quad+\bar{\psi_{k}^{c}} (-1)^k\bigg[A_5\sqrt{2J+1-k}\delta_{k+l,\,2J+1}\delta_{J^\prime,\,J-1/2}
+B_5\sqrt{k}\delta_{k+l,\,2J+2}\delta_{J^\prime,\,J+1/2}
\bigg]\lambda_{l}H_2
\nl
&\quad-\bar{\psi_{k}^{c}} (-1)^k\bigg[-A_5\sqrt{k-1}\delta_{k+l,\,2J+2}\delta_{J^\prime,\,J-1/2}
+B_5\sqrt{2J+2-k}\delta_{k+l,\,2J+3}\delta_{J^\prime,\,J+1/2}
\bigg]\lambda_{l}H_1
\nl
&\quad+\bar{\psi_{k}^{c}} (-1)^k\bigg[A_5^\prime\sqrt{2J+1-k}\delta_{k+l,\,2J+1}\delta_{J^\prime,\,J-1/2}
+B_5^\prime\sqrt{k}\delta_{k+l,\,2J+2}\delta_{J^\prime,\,J+1/2}
\bigg]\gamma^5\lambda_{l}H_2
\nl
&\quad-\bar{\psi_{k}^{c}} (-1)^k\bigg[-A_5^\prime\sqrt{k-1}\delta_{k+l,\,2J+2}\delta_{J^\prime,\,J-1/2}
+B_5^\prime\sqrt{2J+2-k}\delta_{k+l,\,2J+3}\delta_{J^\prime,\,J+1/2}
\bigg]\gamma^5\lambda_{l}H_1
\Bigg\}
\nl
&\quad+\mathrm{h.c.\, terms}\,,
\label{12UVtheory}
\end{align}
where $\lambda$ is a Dirac fermion to construct $\chi^\prime$, similar to the role of $\psi$ for $\chi$, $A_i$, $A^\prime_i$ and $B_i$, $B^\prime_i$ are coupling constants in the UV theory.
We match  the UV theory Eq.\,(\ref{12UVtheory}) onto the effective theory Eq.\,(\ref{12H}), and determine the effective theory coefficients from the UV couplings.

Let us do the matching for the operator $\bar{\chi}_\alpha^m\chi_\beta^l H_i^\dagger H_j$, where $\alpha,\, \beta=1,\,2$ are indices for two Majorana fermions 
$\chi_1$ or $\chi_2$, $m,\,l=1,\,2$ are indices for two components of each Majorana fermion $\chi_\alpha$, and $i,\, j=1,\, 2$ are indices for the two components of Higgs doublet.
The first diagram on the L.H.S with exchange of two $W$ fields yields a group factor
\be
\left(\tilde{T}^a\tilde{T}^b\right)_{\alpha\beta}^{ml}\left(\tau^a \tau^b+\tau^b \tau^a\right)_{ij}=\frac{1}{2}J(J+1)\delta_{\alpha\beta}\delta_{ml}\delta_{ij}\,,
\ee

When the first diagram of L.H.S contains one $W$ and one $B$ exchange, it gives a group factor
\ba
&&\left[\left(\tilde{T}^a\tilde{T}^0\right)^{ml}_{\alpha\beta}+\left(\tilde{T}^0\tilde{T}^a\right)^{ml}_{\alpha\beta}\right]
\left(\tau^a \tau^0+\tau^0 \tau^a\right)_{ij}\nn\\
&&=Y\delta_{\alpha\beta}(J_{ml}^1\sigma_{ij}^1+J_{ml}^3\sigma_{ij}^3)
+iY\sigma_{ij}^2J_{ml}^2\left(-\delta_{\alpha 1}\delta_{\beta 2}+\delta_{\alpha 2}\delta_{\beta 1}\right)\,,
\ea
where $a,\, b=1,\,2,\,3$.

When the first diagram of L.H.S contains two $B$ fields exchange, it gives a group factor 
\ba
\left(\tilde{T}^0\tilde{T}^0\right)_{\alpha\beta}^{ml}\left(\tau^0 \tau^0+\tau^0 \tau^0\right)_{ij}=\frac{Y^2}{2}\delta_{\alpha\beta}\delta_{ml}\delta_{ij}\,.
\ea

Thus, working out all the Feynman rules and the  first diagram of L.H.S contributes to the operator $\bar{\chi}_\alpha^m\chi_\beta^l H_i^\dagger H_j$ a coefficient
\ba
&&-\left[\frac{1}{2}J(J+1)g_2^4\delta_{ml}\delta_{ij}+Yg_1^2g_2^2(J_{ml}^1\sigma_{ij}^1+J_{ml}^3\sigma_{ij}^3)+\frac{Y^2}{2}g_1^4\delta_{ml}\delta_{ij}\right]
\delta_{\alpha\beta}I_{\mathrm{loop}}\nn\\
&&-iYg_1^2g_2^2\sigma_{ij}^2J_{ml}^2\left(-\delta_{\alpha 1}\delta_{\beta 2}+\delta_{\alpha 2}\delta_{\beta 1}\right)I_{\mathrm{loop}}\,,
\label{gaugelhs}
\ea
where $I_{\mathrm{loop}}$ is the loop integral for the first diagram on the L.H.S and
\ba
I_{\mathrm{loop}}=\int \frac{d^d p}{(2\pi)^d}\frac{\gamma^\mu(\slashed{p}+\slashed{q}+M)\gamma_\mu}
{\left[(p+q)^2-M^2+i0\right](p^2+i0)^2}
=\frac{i}{(4\pi)^{2-\epsilon}}\frac{\Gamma(1+\epsilon)}{M^{1+2\epsilon}}(3-2\epsilon)\,,
\ea
with $d=4-2\epsilon$.

The second diagram vanishes by straightforward computation in dimensional regularization.
The only surviving diagram on the R.H.S is the last diagram and other diagrams vanish since the loop integrals are scaleless but dimensionful.

Specifically, let us consider the matching when $J^\prime=J-1/2$ and $Y^\prime=Y-1/2=-Y+1/2=0$. 
Other choices will bring us similar results and the heavy-limit result is the same as we will see in Eq.\,(\ref{evencH}).
When we choose $\alpha=\beta=1$, $i=j=1$ and $m=l$, we obtain
\begin{align}
\mathrm{R.H.S}&=-\frac{i}{2M}\left[\mathbf{Re}(c_{1,1/2})-\mathbf{Re}(c_{2,1/2})(l-J-1)\right]\delta_{ml}\,,
\nl
\mathrm{L.H.S}&=-\left[\frac{1}{2}J(J+1)g_2^4\delta_{ml}\delta_{ij}+Yg_1^2g_2^2(J_{ml}^1\sigma_{ij}^1+J_{ml}^3\sigma_{ij}^3)+\frac{Y^2}{2}g_1^4\delta_{ml}\delta_{ij}\right]I_{\mathrm{loop}}
\nl
&\quad+\frac{i}{4M^\prime}\left(|A_1|^2+|A_3|^2+|A_7|^2+|A_8|^2\right)(2J+1-l)\delta_{ml}\,,
\end{align}
and
\begin{align}
\mathbf{Re}(c_{1,1/2})&=\frac{3}{(4\pi)^2}\left[J(J+1)g_2^4+Y^2 g_1^4\right]-\frac{M}{2M^\prime}\left(|A_1|^2+|A_3|^2+|A_7|^2+|A_8|^2\right)J\,,
\nl
\mathbf{Re}(c_{2,1/2})&=\frac{6g_1^2 g_2^2}{(4\pi)^2}Y-\frac{M}{2M^\prime}\left(|A_1|^2+|A_3|^2+|A_7|^2+|A_8|^2\right)\,.
\end{align}

Then we have the coefficients in Eq.\,(\ref{ABCexprs}),
\begin{align}
A&=\frac{1}{2}\left[\mathbf{Re}(c_{1,1/2})-\mathbf{Re}(c_{2,1/2})(J+1)\right]
\nl
&=\frac{3}{2(4\pi)^2}\left[g_2^4J(J+1)-2g_1^2g_2^2Y(J+1)+Y^2g_1^4\right]+\frac{M}{4M^\prime}\left(|A_1|^2+|A_3|^2+|A_7|^2+|A_8|^2\right)\,,
\nl
B&=\frac{1}{2}\mathbf{Re}(c_{2,1/2})=\frac{3g_1^2 g_2^2}{(4\pi)^2}Y-\frac{M}{4M^\prime}\left(|A_1|^2+|A_3|^2+|A_7|^2+|A_8|^2\right)\,.
\end{align}

To obtain the coefficient $C$ in Eq.\,(\ref{ABCexprs}), let us consider the matching for operator $\bar{\chi}_\alpha^m\chi_\beta^l H_i H_j$ and we choose 
$\alpha=2$, $\beta=1$, $i=j=1$ and $m+l=2J+1$,
\begin{align}
\mathrm{R.H.S}&=-\frac{c_{4,1/2}+c_{3,1/2}^*}{4M}(-1)^{l}\sqrt{l(2J+1-l)}\delta_{l+m,\,2J+1}\,,
\nl
\mathrm{L.H.S}&=\frac{i}{4M^\prime}A_3A_8^*(-1)^l\sqrt{l(2J+1-l)}\delta_{l+m,\,2J+1}\,,
\end{align}
yielding
\ba
C=\frac{c_{4,1/2}+c_{3,1/2}^*}{2}=-i\frac{A_3A_8^*}{2}\frac{M}{M^\prime}\,.
\ea

\bibliography{gen}

\end{document}